\documentclass[twocolumn,showpacs,pra,amsfonts,amsmath,amssymb,floatfix]{revtex4} 
\usepackage{color}
\usepackage{hhline}
\usepackage{mathrsfs}
\usepackage{graphicx}
\usepackage{dcolumn}
\usepackage{bm}
\usepackage{multirow}
\usepackage{booktabs}
\usepackage{afterpage}
\usepackage{amsmath}

\begin{document}

\title{Fully self-consistent optimization of the Jastrow-Slater-type wave function using a similarity-transformed Hamiltonian}

\author{Masayuki Ochi}
\affiliation{Department of Physics, Osaka University, Machikaneyama-cho, Toyonaka, Osaka 560-0043, Japan}
\affiliation{Forefront Research Center, Osaka University, Machikaneyama-cho, Toyonaka, Osaka 560-0043, Japan}

\date{\today}
\begin{abstract}
It has been well established that the Jastrow correlation factor can effectively capture the electron correlation effects, and thus, the efficient optimization of the many-body wave function including the Jastrow correlation factor is of great importance.
For this purpose, the transcorrelated $+$ variational Monte Carlo (TC$+$VMC) method is one of the promising methods, where the one-electron orbitals in the Slater determinant and the Jastrow factor are self-consistently optimized in the TC and VMC methods, respectively.
In particular, the TC method is based on similarity-transformation of the Hamitonian by the Jastrow factor, which enables the efficient optimization of the one-electron orbitals under the effective interaction.
Through test calculations of some closed-shell atoms, He, Be, and Ne, we find that the total energy is in many cases systematically improved by using better Jastrow functions.
We find that even a one-shot TC$+$VMC calculation, where the Jastrow parameters are optimized at the Hartree-Fock$+$VMC level, can yield partial benefits from orbital optimization. It is also suggested that one-shot TC$+$VMC can be a good alternative way for complex systems.
Our study provides important insights for optimizing many-body wave function including the Jastrow correlation factor, which would be of great help for development of highly accurate electronic structure calculations.
\end{abstract}

\maketitle

\section{Introduction}
Accurate description of the electronic structure 
relies on whether the theory captures essential aspects of the electron correlation effects.
One important aspect of many-electron correlation originates from the singular behavior of the Coulomb interaction at electron coalescence points,
which is represented with the Kato cusp condition~\cite{cusp,cusp2}.
From this viewpoint, an explicit inclusion of the electron-electron distance $r_{12}$ into the theory is crucial, as pointed out in pioneering works by Hylleraas~\cite{Hyl1,Hyl2,Hyl3}.
It is now well known that, a slow convergence with respect to the basis-set size for one-electron orbitals is effectively improved by the R12~\cite{R12} and F12~\cite{F12} theories.

Because the R12 and F12 theories aim to an efficient approach to the complete-basis-set (CBS) limit, the CBS limit itself is unchanged by the use of $r_{12}$-dependent functions.
Another attempting idea is to improve the accuracy of the wave function by including the two-electron variables into the wave function.
A representative example is the Jastrow-Slater-type wave function, where the Jastrow factor, which depends on $r_{12}$ and often more complicated coordinates, is multiplied by the Slater determinant.
Quantum Monte Carlo (QMC) methods, such as the variational Monte Carlo (VMC) and diffusion Monte Carlo methods~\cite{QMC}, successfully handle this kind of wave functions not only for atoms and molecules but also for condensed matters~\cite{QMCsolids}.
A key point here is that evaluation of physical quantities for the many-body wave function including the Jastrow factor requires a 3$N$-dimensional integration ($N$: the number of electrons), which is efficiently performed with the Monte Carlo technique.

An alternative way to handle the Jastrow factor was introduced by Boys and Handy~\cite{BoysHandy,Handy}: they proposed to use the Hamiltonian similarity-transformed by the Jastrow factor. 
Then, orbital optimization for the Jastrow-Slater-type wave function is regarded as the Hartree-Fock (HF)  approximation to the similarity-transformed Hamiltonian, which does not require a huge-dimensional integration mentioned above. This is called the transcorrelated (TC) method~\cite{BoysHandy,Handy,Ten-no1,Ten-no2,Umezawa}.
An important advantage of the TC method is that one can apply sophisticated post-HF methods to the similarity-transformed Hamiltonian. To say, the TC method can be a promising starting point of the wave-function theory instead of HF; electron correlation effects are partially taken into account already at the first hierarchy level with respect to the similarity-transformed Hamiltonian.
In fact, previous studies performed TC calculations combined with the coupled-cluster theory~\cite{Ten-no3, TCCC2021},
M{\o}ller-Plesset (MP) perturbation theory~\cite{Ten-no1,Ten-no2},
and configuration interaction (CI) theory~\cite{Umezawa_CIS, LuoVTC, Luo_multiconf, Giner_He, SCI} for atomic and molecular systems.
Canonical TC method~\cite{CanonicalTC} is an important development, in which one employs the idea of the similarity transformation in the context of the F12 theory.
Recently, the canonical TC method was used in conjunction with the variational quantum eigensolver method~\cite{CanonicalTC_qCCSD}.
Also, it is remarkable that the canonical TC method efficiently reduces the number of required Slater determinants in the full-CI (FCI) QMC calculation~\cite{FCI_canoTC_DMRG,FCI_canoTC}.
Here, similarity-transformed FCIQMC, the combination of the TC and FCIQMC methods, has recently been paid much attention~\cite{FCI_TC_elgas, FCI_canoTC_DMRG, FCI_canoTC, FCI_Bedimer, FCI_largeCI, FCI_Hubbard, FCI_1dgas, FCI_cold}.
A recent study that combines the TC method with the quantum computational method is also promising~\cite{McArdle, quantum_simulation}.
Moreover, an efficient treatment of the correlation effects enables one to apply the TC method to solids~\cite{Sakuma,TCaccel,TCjfo,TCPW,TCZnO,TCPP}, including its combination with the CI-singles~\cite{TCCIS} and the second-order MP perturbation theory~\cite{TCMP2}.
The TC and related methods were also applied to the Hubbard model~\cite{TCHubbard,FCI_Hubbard,TCDMRG,LieAlgebra}, 
electron gas~\cite{elgas_Armour,Umezawa_elgas,Sakuma,Luo_elgas,FCI_TC_elgas, perturbation_elgas}, one-dimensional quantum gas with contact interactions~\cite{FCI_1dgas}, and ultracold atoms~\cite{FCI_cold}.

Several post-HF methods focus on improvement over the single Slater determinant, e.g., by using a linear combination of many determinants (CI) and taking the perturbative correction into account (MP perturbation theory).
On the other hand, optimizing the Jastrow factor is another important way to improve the quality of the many-body wave function.
While a parameter-free (i.e. no degree of freedom for optimization) Jastrow function sometimes works well (e.g.~\cite{Ten-no1,Giner_He,Sakuma,Umezawa_elgas}),
parameters in the Jastrow factor have been successfully optimized in the TC method~\cite{Umezawa, Umezawa_beta, LuoTC, LuoVTC, Luo_multiconf, Luo_elgas, TCjfo}.
However, Jastrow optimization in the TC method still has not been well investigated: how it affects the accuracy of the total energy and the pseudoenergy, which is an expectation value of the similarity-transformed Hamiltonian, what is the efficient way for optimization, how the effective interactions in the similarity-transformed Hamiltonian are altered by Jastrow optimization, and so on.
It is noteworthy that recent studies shed light on the importance of the Jastrow optimization in the TC method~\cite{TCatoms_HFJastrow, TCatoms_oneparam, TCatoms_2023}.
While their study uses the Jastrow parameters optimized for the HF orbitals, fully self-consistent optimization of both the Jastrow factor and the Slater determinant is desired.
The full optimization of the Jastrow-Slater-type wave function is also challenging for QMC calculation~\cite{Umrigar_fullopt,Umrigar_fullopt2}.

In this study, we systematically investigate the fully self-consistent solution of the TC method, where both the Jastrow factor and the Slater determinant are optimized.
We use the TC+VMC method~\cite{Umezawa} because VMC is a well-established method for optimizing the Jastrow factor in the Jastrow-Slater-type wave function,
using more sophisticated Jastrow functions compared to the previous study of the TC+VMC method~\cite{Umezawa}.
We investigate both the TC and biorthogonal TC (BITC) methods, where the left and right Slater determinants can be different.
Although many previous studies investigate one of these two formulations, we find that the pseudoenergy is much different between two formulations.
To understand the nature of the similarity transformation, we also investigate how the effective interactions in the similarity-transformed Hamiltonian are altered by Jastrow optimization.
We perform these calculations on some closed-shell atoms, He, Be, and Ne.
We take a sufficiently large number of basis functions for one-electron orbitals to eliminate the basis-set error and see how the correlation effects are described in the optimized Jastrow factor.
We find that the total energy is in many cases systematically improved by using better Jastrow functions. By improving the Jastrow function, the expectation value of the non-Hermitian TC Hamiltonian (pseudoenergy) gets closer to that of the original Hamiltonian for a helium atom. These different estimates of the total energy roughly coincide when one includes the electron-electron-nucleus terms into the Jastrow function.
We also find that one can partially receive a benefit of the orbital optimization even by one-shot TC$+$VMC, where the Jastrow parameters are optimized at the HF$+$VMC level. However, such a superiority of the fully-self-consistent TC$+$VMC calculation does not take place when our simple electron-electron-nucleus Jastrow function is used for Be and Ne atoms,
suggesting that an alternate repetition of TC and VMC optimizations using different guiding principles sometimes causes a difficulty in reaching an accurate solution.
In such a case, one-shot TC$+$VMC might be a good alternative way to apply the TC orbital optimization to the Jastrow-Slater-type wave function.
This study offers a great clue to know how one can efficiently improve a quality of many-body wave function including the Jastrow correlation factor, which would be of great help for development of highly accurate electronic structure calculation.

The paper is organized as follows. In Sec.~\ref{sec:TC}, we briefly present a theoretical framework of the (single-determinant) TC method.
Our implementation of the all-electron TC calculation for the closed-shell atom is presented in Sec.~\ref{sec:TCall}.
In Sec.~\ref{sec:TCVMC}, we introduce the TC$+$VMC method and the Jastrow functions used in this study.
Calculation results are shown in Sec.~\ref{sec:Res}. Section~\ref{sec:Con} is devoted to the conclusion of this study.
Hartree atomic units (a.u.) are used throughout this paper: $\hbar = |e| = m_e = 4\pi\varepsilon_0=1$.

\section{TC method\label{sec:TC}}

Here we briefly describe a theoretical framework of the TC method, a detail of which was presented in previous papers~\cite{BoysHandy, Ten-no1, Umezawa}. Hamiltonian $\mathcal{H}$ for an $N$-electron system under an external potential $v_{\mathrm{ext}}(\mathbf{r})$ is written as,
\begin{equation}
\mathcal{H}=\sum_{i=1}^{N} \left( -\frac{1}{2}\nabla_i ^2 + v_{\mathrm{ext}}(\mathbf{r}_i) \right) +
\sum_{i=1}^{N}\sum_{j>i}^N \frac{1}{|\mathbf{r}_i-\mathbf{r}_j|}.\label{eq:Hamil}
\end{equation} 
First, we formally factorize the many-body wave function $\Psi$ as $\Psi=F\Phi$ where $F$ is the Jastrow factor,
\begin{equation}
F=\mathrm{exp}(-\sum_{i,j(>i)}^N u(x_i,x_j)), \label{eq:Jastrowfactor}
\end{equation}
and $\Phi$ is defined as $\Phi \equiv \Psi/F$. 
Here, $x=(\mathbf{r},\sigma)$ denotes a set of spatial and spin coordinates associated with an electron.
Here we assume the Jastrow function $u(x_i,x_j)$ to be symmetric, i.e., $u(x_i,x_j)=u(x_j,x_i)$, without loss of generality.
Next, we introduce a similarity-transformed Hamiltonian,
\begin{equation}
\mathcal{H}_{\mathrm{TC}} \equiv F^{-1}\mathcal{H}F,
\end{equation}
by which the Schr{\"o}dinger equation is rewritten as,
\begin{equation}
\mathcal{H}\Psi = E\Psi \Leftrightarrow \mathcal{H}_{\mathrm{TC}}\Phi = E \Phi \label{eq:simtr}.
\end{equation}
In this way, electron correlation effects described with the Jastrow factor are incorporated into
the similarity-transformed Hamiltonian $\mathcal{H}_{\mathrm{TC}}$, which we called the TC Hamiltonian hereafter.
$\mathcal{H}_{\mathrm{TC}}$ can be explicitly written as,
\begin{gather}
\mathcal{H}_{\mathrm{TC}}=\sum_{i=1}^{N} \left( -\frac{1}{2}\nabla_i ^2 + v_{\mathrm{ext}}(\mathbf{r}_i) \right) +
\sum_{i=1}^{N}\sum_{j>i}^N v_{\mathrm{2body}}(x_1,x_2)  \notag \\
- \sum_{i=1}^{N}\sum_{j>i}^N \sum_{k>j}^N v_{\mathrm{3body}}(x_1,x_2,x_3),
\end{gather}
where $v_{\mathrm{2body}}(x_1,x_2)$ and $v_{\mathrm{3body}}(x_1,x_2,x_3)$ are the effective interactions in the TC Hamiltonian defined as,
\begin{gather}
v_{\mathrm{2body}}(x_1,x_2)\notag\\
\equiv \frac{1}{|\mathbf{r}_1-\mathbf{r}_2|}+\frac{1}{2}\bigg[ \nabla_1^2 u(x_1,x_2)+\nabla_2^2 u(x_1,x_2)\notag \\
-(\nabla_1 u(x_1,x_2))^2-(\nabla_2 u(x_1,x_2))^2\bigg] \notag \\
+ \nabla_1 u(x_1,x_2)\cdot \nabla_1 + \nabla_2 u(x_1,x_2)\cdot \nabla_2, \label{eq:V2body}
\end{gather}
and
\begin{gather}
v_{\mathrm{3body}}(x_1,x_2,x_3)\notag\\
\equiv\nabla_1 u(x_1,x_2)\cdot \nabla_1 u(x_1,x_3) 
+ \nabla_2 u(x_2,x_1) \cdot \nabla_2 u(x_2,x_3) \notag \\
+ \nabla_3 u(x_3,x_1) \cdot \nabla_3 u(x_3,x_2).  \label{eq:V3body}
\end{gather}
It is characteristic that the TC Hamiltonian does not include four- or higher-body effective interactions as long as a Jastrow factor including only up to two-electron correlations is used.

By approximating $\Phi$ to be a single Slater determinant consisting of one-electron orbitals: $\Phi=\mathrm{det}[ \phi_i(x_j) ]$, the following one-body self-consistent-field (SCF) equation is derived (see, e.g., \cite{Umezawa}):
\begin{align}
\left( -\frac{1}{2}\nabla_1^2 +v_{\mathrm{ext}}(\mathbf{r}_1) \right) \phi_i (\mathbf{r}_1)\notag \\
+ \sum_{j=1}^N
\int \mathrm{d}\mathbf{r}_2\  \phi_j^*(\mathbf{r}_2) v_{\mathrm{2body}}(x_1,x_2)
\mathrm{det} \left[
\begin{array}{rrr}
\phi_i(\mathbf{r}_1) & \phi_i(\mathbf{r}_2) \\
\phi_j(\mathbf{r}_1) & \phi_j(\mathbf{r}_2) \\
\end{array} \right] \notag \\
- \sum_{j=1}^N \sum_{k>j}^N
\int \mathrm{d}\mathbf{r}_2 \mathrm{d}\mathbf{r}_3\  \phi_j^*(\mathbf{r}_2)\phi_k^*(\mathbf{r}_3)v_{\mathrm{3body}}(x_1,x_2,x_3)  \notag \\
\times 
\mathrm{det} \left[
\begin{array}{rrr}
\phi_i(\mathbf{r}_1) & \phi_i(\mathbf{r}_2) &  \phi_i(\mathbf{r}_3) \\
\phi_j(\mathbf{r}_1) & \phi_j(\mathbf{r}_2) & \phi_j(\mathbf{r}_3) \\
\phi_k(\mathbf{r}_1) & \phi_k(\mathbf{r}_2) & \phi_k(\mathbf{r}_3)
\end{array} \right]
= \sum_{j=1}^N \epsilon_{ij} \phi_j(\mathbf{r}_1), \label{eq:SCF}
\end{align}
where the orthonormal condition, $\langle \phi_i | \phi_j \rangle = \delta_{i,j}$, is imposed.
The TC one-electron orbitals $\phi_i(\mathbf{r})$ are optimized by solving Eq.~(\ref{eq:SCF}).
$\Phi$ can be systematically improved over a single Slater determinant by applying the post-HF theories to the similarity-transformed Hamiltonian~\cite{Ten-no1,Ten-no2,Ten-no3,TCCIS,TCMP2}, which is an important advantage of the TC method.
This feature is validated by the equivalency between the two eigenvalue problems presented in Eq.~(\ref{eq:simtr}): to say, the exact eigenstate of the original Hamiltonian $\mathcal{H}$ can be immediately constructed from the exact eigenstate of $\mathcal{H}_{\mathrm{TC}}$ by $\Psi=F\Phi$.
In this study, however, we concentrate on the case when $\Phi$ is approximated as a single Slater determinant.

The total energy,
\begin{equation}
E = \frac{\langle \Psi | \mathcal{H} | \Psi \rangle}{\langle \Psi  | \Psi \rangle} \label{eq:totE},
\end{equation}
equals to the TC pseudoenergy,
\begin{equation}
E_{\mathrm{TC}} = \mathrm{Re} \bigg[ \frac{\langle \Phi | \mathcal{H}_{\mathrm{TC}} | \Phi \rangle}{\langle \Phi  | \Phi \rangle} \bigg],\label{eq:EtotTC}
\end{equation}
when $\Phi$ is the exact eigenstate of $\mathcal{H}_{\mathrm{TC}}$. While this is of course not true for an approximate $\Phi$, we can still approximately evaluate the total energy by $E_{\mathrm{TC}}$, which does not require $N$-dimensional integration.
For example, when $\Phi$ is a single Slater determinant, $E_{\mathrm{TC}}$ can be calculated by 9-dimensional (three-body) integration.
This is another great advantage of the TC method from the viewpoint of computational cost.
However, accuracy of the approximation $E_{\mathrm{TC}} \simeq E$ should be carefully checked, which is one of the main objectives of this study.

We also mention the biorthogonal formulation of the TC method, which we called the BITC method.
The BITC method was applied to molecules~\cite{Ten-no2} and recently also to solids~\cite{TCMP2}.
A detailed description of the BITC method can be found in these literatures.
In the BITC method, we use left and right Slater determinants consisting of different one-electron orbitals: $X=\mathrm{det}[\chi_i(x_j)]$ and $\Phi=\mathrm{det}[\phi_i(x_j)]$, respectively, with the biorthogonal condition $\langle \chi_i | \phi_j \rangle = \delta_{i,j}$ and the normalization condition $\langle \phi_i | \phi_i \rangle = 1$. Then a one-body SCF equation becomes slightly different from Eq.~(\ref{eq:SCF}) as follows,
\begin{align}
\left( -\frac{1}{2}\nabla_1^2 +v_{\mathrm{ext}}(\mathbf{r}_1) \right) \phi_i (\mathbf{r}_1)\notag \\
+ \sum_{j=1}^N
\int \mathrm{d}\mathbf{r}_2\  \chi_j^*(\mathbf{r}_2) v_{\mathrm{2body}}(x_1,x_2)
\mathrm{det} \left[
\begin{array}{rrr}
\phi_i(\mathbf{r}_1) & \phi_i(\mathbf{r}_2) \\
\phi_j(\mathbf{r}_1) & \phi_j(\mathbf{r}_2) \\
\end{array} \right] \notag \\
- \sum_{j=1}^N \sum_{k>j}^N
\int \mathrm{d}\mathbf{r}_2 \mathrm{d}\mathbf{r}_3\  \chi_j^*(\mathbf{r}_2)\chi_k^*(\mathbf{r}_3)v_{\mathrm{3body}}(x_1,x_2,x_3)  \notag \\
\times 
\mathrm{det} \left[
\begin{array}{rrr}
\phi_i(\mathbf{r}_1) & \phi_i(\mathbf{r}_2) &  \phi_i(\mathbf{r}_3) \\
\phi_j(\mathbf{r}_1) & \phi_j(\mathbf{r}_2) & \phi_j(\mathbf{r}_3) \\
\phi_k(\mathbf{r}_1) & \phi_k(\mathbf{r}_2) & \phi_k(\mathbf{r}_3)
\end{array} \right]
= \epsilon_{ii} \phi_i(\mathbf{r}_1). \label{eq:SCF_BITC}
\end{align}
By rewriting Eq.~(\ref{eq:SCF_BITC}) as $\hat{h}\phi_i = \epsilon_{ii}\phi_i$, we can solve this equation by diagonalization the operator $\hat{h}$. Since the one-body SCF equation for the left orbital $\chi$ is $\hat{h}^{\dag} \chi_i = \epsilon^*_{ii}\chi$, we can simultaneously get $\chi$ as the left eigenstates of $\hat{h}$ in the above diagonalization.
The BITC pseudoenergy is defined as,
\begin{equation}
E_{\mathrm{BITC}} = \mathrm{Re} \bigg[  \frac{\langle X | \mathcal{H}_{\mathrm{TC}} | \Phi \rangle}{\langle X  | \Phi \rangle} \bigg].\label{eq:EtotBITC}
\end{equation}
Because the similarity transformation of Hamiltonian introduces non-Hermiticity, such formulation yields a different result from the ordinary TC method presented above.
A difference between the TC and BITC methods is also an issue that we shall investigate in this paper.

\section{All-electron TC calculation for an atom\label{sec:TCall}}

Here we describe how we perform the all-electron calculation of the TC method.
In this study, we focus on the non-relativistic treatment of the closed-shell atom with a nucleus charge $Z$ placed at the origin.

\subsection{One-electron orbitals and a basis set\label{sec:orb_orthognal}}

We represent the left and right one-electron orbitals by the spherical harmonics $Y_{l_i,m_i}$ and the radial functions as follows:
\begin{eqnarray}
\phi_{n_i,l_i,m_i,\sigma_i}(\mathbf{r})=Y_{l_i m_i}(\Omega)\frac{\phi^{\mathrm{rad}}_{n_i,l_i,\sigma_i}(r)}{r}, \\
\chi_{n_i,l_i,m_i,\sigma_i}(\mathbf{r})=Y_{l_i m_i}(\Omega')\frac{\chi^{\mathrm{rad}}_{n_i,l_i,\sigma_i}(r)}{r},
\end{eqnarray}
where $n_i, l_i, m_i, \sigma_i$ denote the principal, azimuthal, magnetic, and spin quantum numbers, respectively.
We often abbreviate $\phi^{\mathrm{rad}}_{n_i,l_i,\sigma_i}$ as $\phi^{\mathrm{rad}}_i$ hereafter.
We expand the radial functions with the azimuthal quantum number $l_i$
using a following basis function (see, e.g. \cite{Laguerre1,Laguerre2}):
\begin{equation}
f_n^{l_i}(r) = (2\alpha)^{l_i+\frac{3}{2}} \sqrt{\frac{n!}{(n+2l_i+2)!}} r^{l_i+1} L_n^{(2l_i+2)}(2\alpha r) e^{-\alpha r}, \label{eq:basis}
\end{equation}
where $\alpha = \sqrt{-2\epsilon_{\mathrm{HO}}}$ is a scaling factor using the eigenenergy of the highest occupied orbitals, $\epsilon_{\mathrm{HO}}$, and
 $L_n^{(k)}$ is an associated Laguerre polynomial: 
 \begin{equation}
 L_n^{(k)}(x) \equiv \frac{e^x x^{-k}}{n!}\frac{\mathrm{d}^n}{\mathrm{d}x^n} (e^{-x}x^{n+k}),
 \end{equation}
some examples of which are $L_0^{(k)}(x)=1, L_1^{(k)}(x)=-x+k+1$, and so on.
 This basis function satisfies correct asymptotic behaviors of the radial function~\cite{asympt}: $f_n^{l_i}(r) \to e^{-\alpha r}$ for $r\to \infty$ and $f_n^{l_i}(r) \to r^{l_i+1}$ for $r\to 0$. Hence, any orbitals that can be represented by a product of $r^{l_i+1}e^{-\alpha r}$ and a polynomial of $r$ can be expanded with this basis set. Orthonormality of this basis set is readily verified (see Appendix A).
 Note that the scaling factor $\alpha$ is updated in each SCF loop because it depends on $\epsilon_{\mathrm{HO}}$~\cite{note_homo}.

\subsection{TC-SCF equation for the radial function}

The TC-SCF equation for the radial function of the right orbitals reads
\begin{equation}
\left[ -\frac{1}{2}\frac{\mathrm{d}^2}{\mathrm{d}r^2}+\frac{l_i(l_i+1)}{2r^2}+\hat{V}[\phi,\chi]\right] \phi_i^{\mathrm{rad}}(r) = \sum_{j}^N \epsilon_{ij} \phi_j^{\mathrm{rad}}(r),\label{eq:AEeq}
\end{equation}
where $\hat{V}$ denotes the one-, two-, and three-body potentials as described in more detail later in this paper.
Equations for the left orbitals can be obtained in the same way and so are not shown here.

We solved the TC-SCF equation~(\ref{eq:AEeq}) by evaluating the matrix elements of the left-hand side of this equation with respect to the basis functions defined in Eq.~(\ref{eq:basis}), and diagonalizing the matrix~\cite{note_matrix}. 
Since we assume that the one-electron orbital is a product of the spherical harmonics and radial function, diagonalization is separately performed for each $(l,m,\sigma)$.
This procedure is repeated until the self-concistency with respect to the orbitals is achieved, because $\hat{V}[\phi,\chi]$ depends on the orbitals $\phi$ and $\chi$. The matrix elements are evaluated on the real-space grid points.
To describe a rapid oscillation of the wave functions near a nucleus, we used a log mesh $\rho=\ln(r)$ for the real-space grid.
We applied a sufficiently large cutoff for the range of $\rho$ with the boundary condition that $\phi^{\mathrm{rad}}$ and $\chi^{\mathrm{rad}}$ go to zero at the both end points.

Here, we comment on the Gram-Schmidt orthonormalization of the orbitals. Because of the non-Hermitian character of the similarity-transformed Hamiltonian, the eigenstates of the TC-SCF equation are not orthogonalized. Thus, actual calculation is proceeded as follows. First, we diagonalize the TC-SCF matrix and get its eigenstates. Next, we perform Gram-Schmidt orthonormalization of the orbitals to satisfy the orthogonality. Note that the orbitals with different $(l_i,m_i,\sigma_i)$ are orthogonal even when the non-Hermiticity takes place, and hence this orthogonalization is performed within the same $(l_i,m_i,\sigma_i)$.
These orthonormalized orbitals are the TC one-electron orbitals shown in this paper. Because the Gram-Schmidt orthonormalization (i.e. linear combination within the occupied orbitals) does not change the Slater determinant except a constant factor as is guaranteed by the antisymmetry of the determinant, this procedure does not change the many-body wave function. By using the orthonormalized orbitals, we can derived the TC-SCF equation, Eq.(\ref{eq:SCF}). The detail of this procedure was explained in Ref.~\cite{Umezawa}. For the BITC method, the biorthogonal condition is imposed instead of the orthogonal condition, and so this issue does not take place.

As adopted in Ref.~\cite{Umezawa}, we applied the Gram-Schmidt orthonormalization to the orbitals in ascending order of the eigenvalues. In other words, the Gram-Schmidt orthonormalization starts from the eigenstate with the lowest orbital energy. Although the order of the orbitals for the Gram-Schmidt orthonormalization can change the shape of each orbital, the orbital energies and the total energy are not affected by the following reasons. As for the orbital energies, we have proved in Ref.~\cite{TCPP} that the diagonal element of the eigenvalue matrix, $\epsilon_{ii}$, is not changed by the Gram-Schmidt orthonormalization. The total energy is also unaffected by the order because of the invariance of the Slater determinant against the Gram-Schmidt orthonormalization as mentioned in the previous paragraph. Thus, what one-body SCF equations impose, which is formally equivalent to $\delta \Phi/\delta \phi_i^* = 0$ for all the occupied orbitals $i$, are not affected by the order of the Gram-Schmidt orthonormalization. It is an important future issue that how the orbital shapes are changed by the choice of the order of the orbitals in the Gram-Schmidt orthonormalization, while all the results shown in this paper including the $1s$ orbitals of the helium atom, where no Gram-Schmidt orthonormalization takes place, are unaffected by it.

\subsection{One-body terms in the TC-SCF equation\label{sec:onebody}}

A matrix element for the one-body terms in the TC-SCF equation (\ref{eq:AEeq}),
\begin{equation}
\int_0^{\infty} f_m^{l_i}(r) \left[ -\frac{1}{2}\frac{\mathrm{d}^2}{\mathrm{d}r^2} + \frac{l_i(l_i+1)}{2r^2} - \frac{Z}{r} \right] f_n^{l_i}(r)\ \mathrm{d}r \label{eq:1body},
\end{equation}
can be rewritten as follows (see Appendix B):
\begin{widetext}
\begin{equation}
-\frac{\alpha^2}{2} \delta_{m,n} + \left(-Z + \alpha \frac{(l_i+1)(2l_i+2\mathrm{min}(m,n)+3)}{2l_i+3} \right) \int_0^{\infty} f_m^{l_i}(r) \frac{1}{r} f_n^{l_i}(r)\ \mathrm{d}r, \label{eq:1body_2}
\end{equation}
\end{widetext}
and we numerically evaluated the integral $\langle f_m^{l_i} | 1/r | f_n^{l_i} \rangle$.
By using a log mesh, $\mathrm{d}r = r\mathrm{d}\rho$ removes a diverging behavior ($1/r$) of the integrand.

If one includes a one-body Jastrow function, a one-body effective potential will appear in the similarity-transformed Hamiltonian.
In this study, we did not include any one-body Jastrow function in the TC calculation, because this is a duplicated degree of freedom with one-electron orbitals.
To say, a one-body Jastrow function is not necessary when one optimizes one-electron orbitals.

Nevertheless, we here make few notes for including the one-body Jastrow function, because it can be a useful way for some purposes, e.g., for imposing the (nucleus-electron) cusp condition on the one-body Jastrow function instead of the one-electron orbitals.
One-body effective potentials can be easily obtained by substituting the one-body Jastrow function for Eqs.~(\ref{eq:V2body})-(\ref{eq:V3body}).
For using a one-body Jastrow function, one should properly change a basis set of the one-electron orbitals, Eq.~(\ref{eq:basis}), because an asymptotic behavior of the one-electron orbitals can be altered by the Jastrow factor. Regarding this point, we note that the localizing property of the one-electron orbitals near a nucleus can be lost depending on the one-body Jastrow function, which can induce numerical difficulty in solving the TC-SCF equation.
Thus, it will be better to properly restrict the degree of freedom of the one-body Jastrow function even if one would like to include it in the TC calculation.

\subsection{Two- and three-body terms in the TC-SCF equation}

The one-body terms in the TC-SCF equation, Eq.~(\ref{eq:AEeq}), are easily evaluated because they come down to the one-dimensional integral of the smooth function.
However, the two- and three-body terms require higher-dimensional integration, involving the angle coordinates of the orbitals appearing in 
$\hat{V}[\phi,\chi]$.
In this study, we performed Monte Carlo integration for the two- and three-body terms in the TC-SCF equation for simplicity.

The Monte Carlo sampling is performed in the following procedure.
We go back to the SCF equation, Eq.~(\ref{eq:SCF}), and define the two-body matrix element as
\begin{gather}
\langle F^{lm}_{n_1} | v_{\mathrm{2body}} | F^{lm}_{n_2} \rangle \equiv
\sum_{j=1}^N
\int\mathrm{d}\mathbf{r}_1 \mathrm{d}\mathbf{r}_2
(F^{lm}_{n_1}(\mathbf{r}_1))^* \phi_j^*(\mathbf{r}_2) \notag\\
\times v_{\mathrm{2body}}(x_1,x_2)
\mathrm{det} \left[
\begin{array}{rrr}
F^{lm}_{n_2}(\mathbf{r}_1) & F^{lm}_{n_2}(\mathbf{r}_2) \\
\phi_j(\mathbf{r}_1) & \phi_j(\mathbf{r}_2) \\
\end{array} \right], \label{eq:v2body_matrix}
\end{gather}
where
\begin{equation}
F^{l,m}_n(\mathbf{r}) = Y_{lm}(\Omega_1)\frac{f_n^{l}(r_1)}{r_1} 
\end{equation}
is a basis function of the orbitals.
Since we assume that the one-electron orbital is a product of the spherical harmonics and the radial function, we only consider the diagonal matrix element of the quantum numbers $(l,m,\sigma)$, while all occupied $(l_j,m_j,\sigma_j)$ are summed over for the state $j$.
$\langle F^{lm}_{n_1} | v_{\mathrm{2body}} | F^{lm}_{n_2} \rangle$ can be regarded as a two-body part of $\langle f^{l}_{n_1} | \hat{V}[\phi,\chi]| f^{l}_{n_2} \rangle$.
Monte Carlo sampling for Eq.~(\ref{eq:v2body_matrix}) is performed in the ($\mathbf{r}_1, \mathbf{r}_2$) space.
Because we can fix $\phi_2=0$ (an angle coordinate of $\mathbf{r}_2$) by symmetry of the orbitals, five-dimensional integration is required.
Evaluation of the three-body terms can be done in the same way, which requires eight-dimensional integration.

We note that the two-body terms like $\nabla_1 u(x_1, x_2) \cdot \nabla_1 \phi_i (\mathbf{r_1})$ depend on the gradient of the orbital. For handling these terms, we used
\begin{align}
\nabla \phi_i(\mathbf{r}) = \left( -\frac{\phi^{\mathrm{rad}}_{i}(r)}{r^2} + \frac{1}{r}\frac{\mathrm{d} \phi_{i}^{\mathrm{rad}} (r)}{\mathrm{d}r} \right) Y_{l_i m_i}(\Omega) \mathbf{e}_r \nonumber \\
+\frac{\phi^{\mathrm{rad}}_{i}(r)}{r^2} \bigg( \frac{\partial Y_{l_i m_i}(\Omega)}{\partial \theta} \mathbf{e}_{\theta} +  \frac{im_i}{\sin\theta}Y_{l_i m_i}(\Omega) \mathbf{e}_{\varphi}  \bigg) ,\label{eq:nablaphi}
\end{align}
and the following formula (see Appendix C),
\begin{widetext}
\begin{equation}
\frac{\partial Y_{l_i m_i}(\Omega)}{\partial \theta} = sgn(m_i)\sqrt{(l_i-|m_i|)(l_i+|m_i|+1)} Y_{l_i,m_i+sgn(m_i)}(\Omega) e^{-i sgn(m_i) \varphi} + |m_i| \frac{\cos \theta}{\sin \theta} Y_{l_i,m_i}(\Omega), \label{eq:Ylmderiv}
\end{equation}
\end{widetext}
where $sgn(m_i) = +1\ (m_i\geq 0),\ \ -1\ (m_i<0)$, for evaluating the derivative of the spherical harmonics.
For evaluating the derivative of the radial functions in Eq.~(\ref{eq:nablaphi}), we simply adopted the finite-difference method.

We note that the Jacobian $r^2 \sin \theta$ removes the diverging function ($1/r$ and $1/\sin \theta$) in Eq.~(\ref{eq:nablaphi}) and other potential terms, by which numerical difficulty is avoided. Derivative of the Jastrow function in the effective potentials presenting in Eqs.~(\ref{eq:V2body})-(\ref{eq:V3body}) is evaluated analytically (see Appendix D). The Jastrow function used in this study is shown in Section~\ref{sec:jastrow}.

\section{TC$+$VMC method\label{sec:TCVMC}}

By performing the TC calculation presented in Section~\ref{sec:TCall}, one can optimize the one-electron orbitals for a given Jastrow function.
To optimize the Jastrow function for given one-electron orbitals, we performed the VMC calculation.
In VMC calculations, one can minimize the total energy, Eq.~(\ref{eq:totE}), or the variance,
\begin{equation}
\sigma^{2} = \frac{\langle \Psi | (\mathcal{H}-E_{\mathrm{VMC}})^2 | \Psi \rangle}{\langle \Psi  | \Psi \rangle},
\end{equation}
both of which are evaluated with the Monte Carlo integration. In this paper, we do not describe technical details of VMC, and instead refer the readers to a review article~\cite{QMC}.

\begin{figure}
\begin{center}
\includegraphics[width=8.4 cm]{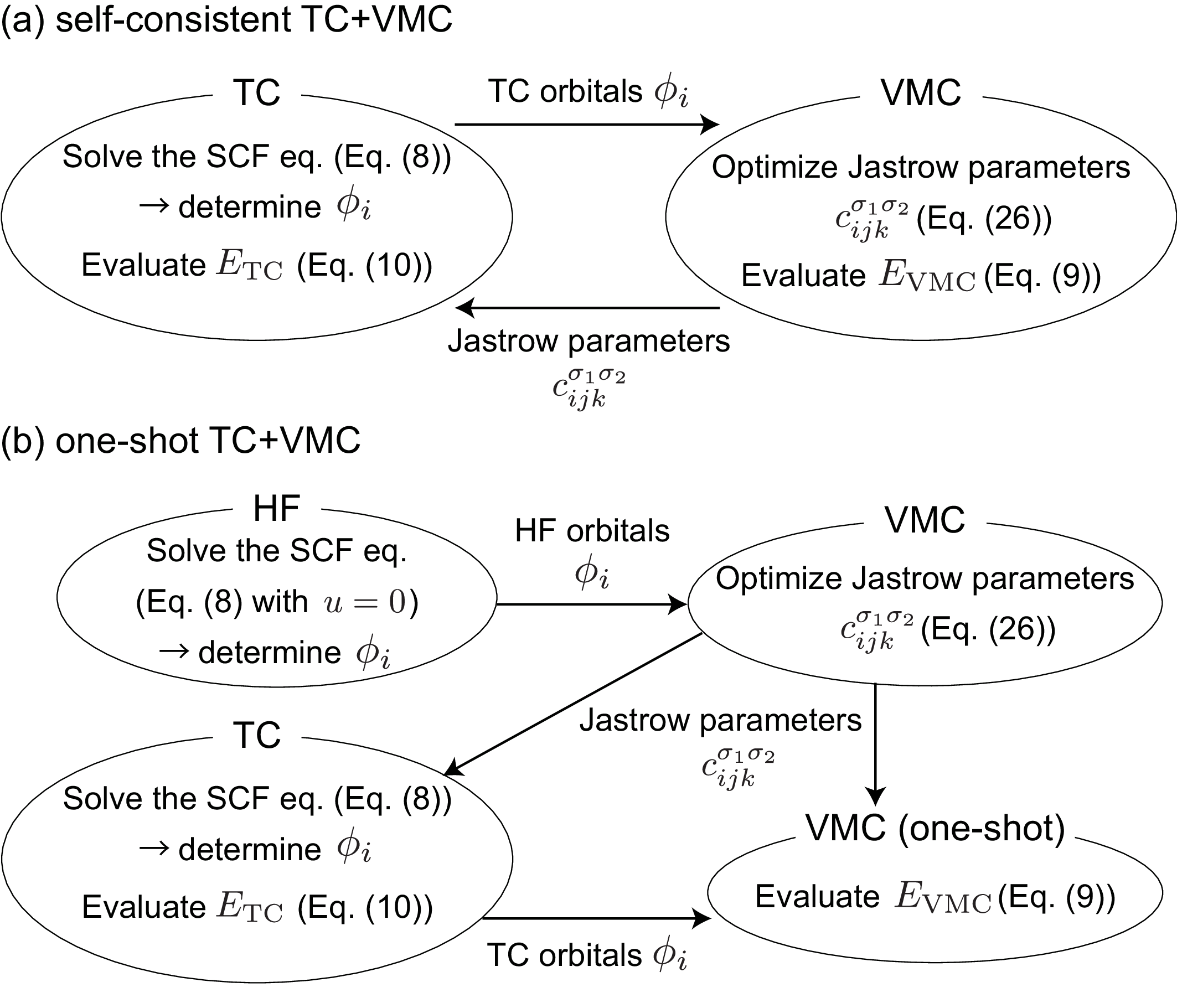}
\caption{Calculation procedure for (a) self-consistent TC$+$VMC and (b) one-shot TC$+$VMC.}
\label{fig:scheme}
\end{center}
\end{figure}

In the TC$+$VMC method, one repeats the TC and VMC calculations alternately for optimizing both the one-electron orbitals and the Jastrow function.
We stopped this iteration when one finds that an additional VMC calculation no longer improves the Jastrow function.
Calculation procedure of self-consistent TC$+$VMC is shown in Fig.~\ref{fig:scheme}(a).
In Fig.~\ref{fig:scheme}, we also present the calculation procedure for one-shot TC$+$VMC, which shall be investigated in Sec.~\ref{sec:oneshot}.

We should mention a small difference between our TC$+$VMC calculation and that studied by Umezawa $et$ $al$.~\cite{Umezawa}
In Ref.~\cite{Umezawa}, the variance defined for the TC Hamiltonian,
\begin{equation}
\sigma^{2}_{\mathrm{TC}} = \frac{\langle \Phi | (\mathcal{H}^{\dag}_{\mathrm{TC}}-E_{\mathrm{TC}}) (\mathcal{H}_{\mathrm{TC}}-E_{\mathrm{TC}}) | \Phi \rangle}{\langle \Phi  | \Phi \rangle},
\end{equation}
is minimized instead of $\sigma^2$. Both guiding principles should work because both of $\sigma^{2}$ and $\sigma^{2}_{\mathrm{TC}}$ become zero for the exact ground state. We shall briefly discuss their difference in Section~\ref{sec:Res}.

We found that TC$+$VMC calculation often fails to converge (i.e., each of TC and VMC gives a converged result while an alternate repetition of TC and VMC calculations does not reach convergence) when the energy is minimized in VMC calculation. In that case, one-electron orbitals become more and more localized near the nucleus by repeating TC and VMC calculations.
In principle, it is not guaranteed that the convergence is achieved by repeating TC and VMC calculations based on different guiding principles for each. Although it is difficult to pin down the cause for the failure in TC$+$VMC (energy minimization), in our experience, variance optimization in VMC works well in TC$+$VMC calculation. Thus, we adopted variance minimization for VMC calculation in this study.
However, we also found that, TC$+$VMC (energy minimization) can reach convergence in some cases by including one-body terms in the Jastrow function for VMC. We shall discuss this issue in Sec.~\ref{sec:one_body_emin}, while VMC (variance minimization) was adopted in other sections.

\subsection{Jastrow functions used in this study\label{sec:jastrow}}

In this study, we used the following Jastrow function:
\begin{equation}
u(x_1,x_2) = \sum_{(i,j,k)\in S} c^{\sigma_1\sigma_2}_{ijk} \bar{r}^i_{12} \bar{r}_1^j \bar{r}_2^k,\label{eq:Jastrow_pade}
\end{equation}
where $S$ is a combination of $(i,j,k)$ considered in this Jastrow function, and $\bar{r}_{12}$ and $\bar{r}_i$ ($i=$ 1,2) are defined as,
\begin{equation}
\bar{r}_{12} = \frac{r_{12}}{r_{12}+a_{12}},\ \ \bar{r}_i = \frac{r_i}{r_i + a}\ \ (i=1,2),
\end{equation}
with $r_{12} = |\mathbf{r}_1 - \mathbf{r}_2|$ and $r_i=|\mathbf{r}_i|$. This Jastrow function is often used in VMC studies~\cite{poly1,poly2}.
For simplicity, we set $a_{12}=a$. 
Since $u$ is a symmetric function, $c^{\sigma_1\sigma_2}_{ijk} =c^{\sigma_2\sigma_1}_{ikj}$ is imposed.
We also assume $c^{\uparrow \uparrow}_{ijk} = c^{\downarrow \downarrow}_{ijk}$ and $c^{\uparrow \downarrow}_{ijk} = c^{\downarrow \uparrow}_{ijk}$ since we consider non-spin-polarized atoms here.
For satisfying the cusp conditions~\cite{cusp,cusp2}, we set
\begin{equation}
c^{\uparrow \uparrow}_{100} = c^{\downarrow \downarrow}_{100} = \frac{a_{12}}{4},\ \ 
c^{\uparrow \downarrow}_{100} = c^{\downarrow \uparrow}_{100} = \frac{a_{12}}{2}.
\end{equation}
and $c^{\sigma_1\sigma_2}_{1jk} = 0$ except $j=k=0$ for simplicity.
Note that this Jastrow function is actually spin-contaminated and does not satisfy the cusp condition. This deficiency can be avoided by constructing the Jastrow factor with the permutation operator, but this procedure introduces non-terminating series of interaction in the TC Hamiltonian~\cite{Ten-nocusp}. Thus, we adopt this approximate cusp condition here. Fortunately, a VMC study reported that an effect of spin contamination on accuracy of the wave function and its energy is small~\cite{cuspUmrigar}.
We also imposed $c^{\sigma_1\sigma_2}_{i1k} = c^{\sigma_1\sigma_2}_{ij1} = 0$ since our one-electron orbitals satisfy the nucleus cusp condition, and $c^{\sigma_1\sigma_2}_{000}=0$ since this component plays no role in improving a quality of the wave function.

In this study, we investigate the following three cases: 
\begin{itemize}
\item $S_{\mathrm{minimal}}=\{(1,0,0)\}$, \item $S_{\mathrm{ee}}=\{ (i,0,0) | 1\leq i \leq 4\}$, \item $S_{\mathrm{een}}=S_{\mathrm{ee}} \cup \{ (0,2,2), (2,2,0), (2,0,2), (2,2,2)\}$.
\end{itemize}
$S_{\mathrm{minimal}}$ represents the minimal Jastrow function satisfying the cusp conditions, which only has a one free parameter $a$.
$S_{\mathrm{ee}}$ corresponds to an electron-electron Jastrow function, and few electron-electron-nucleus terms are added for $S_{\mathrm{een}}$.
In total, $S_{\mathrm{minimal}}$, $S_{\mathrm{ee}}$, and $S_{\mathrm{een}}$ contain zero, six, and twelve independent optimizable parameters (excluding $a$) across spin-pair combinations.
One-body Jastrow functions, $(i,j,k)=(0,j,0)$ and $(0,0,k)$, were not included in this study because these are duplicated degrees of freedom with one-electron orbitals as explained in Section~\ref{sec:onebody}, except in Sec.~\ref{sec:one_body_emin}. In Sec.~\ref{sec:one_body_emin}, we shall see that including one-body Jastrow functions in VMC calculations can improve the convergence of TC$+$VMC calculations where energy minimization was adopted in VMC. In that case, we additionally included the cuspless one-body terms, $(i,j,k)=(0,j,0)$ and $(0,0,k)$ with $2\leq j,k \leq 4$ only in VMC calculations.
\section{Results\label{sec:Res}}

In this paper, we first applied the TC$+$VMC method to a helium atom as a test case.
While a helium atom can be regarded as one of the simplest ``many-body'' systems, it includes essential aspects of the electron correlation effects. We also used the BITC method, as described in the following sections.
We also applied the TC$+$VMC method to beryllium and neon atoms for comparison, as shown in Sec.~\ref{sec:other_atoms}.

\subsection{Computational details\label{sec:Computationa_details}}

For the HF, TC, and BITC calculations, a radial mesh for one-electron orbitals was set in the range of $1.2\times 10^{-4}$ Bohr $\leq r \leq$ $4.0\times 10^{2}$ Bohr for He, $1.2\times 10^{-4}$ Bohr $\leq r \leq$ $1.1\times 10^{3}$ Bohr for Be, and $4.5\times 10^{-5}$ Bohr $\leq r \leq$ $4.0\times 10^{2}$ Bohr for Ne. 
Within this range, a log mesh with mesh points of $N_{\mathrm{mesh}}$ was taken in the HF, TC, and BITC calculations, while $N_{\mathrm{coarse}}$ times coarser grid (i.e., the number of mesh points is $N_{\mathrm{mesh}}/N_{\mathrm{coarse}}$) was used as an input orbital for VMC for efficient Monte Carlo sampling in VMC.
For He and Be,  $N_{\mathrm{mesh}}= 80,000$ and $N_{\mathrm{coarse}}=500$  (i.e., $N_{\mathrm{mesh}}/N_{\mathrm{coarse}}=160$ for VMC) were taken.
For Ne, $N_{\mathrm{mesh}}= 100,800$ and $N_{\mathrm{coarse}}=840$  (i.e., $N_{\mathrm{mesh}}/N_{\mathrm{coarse}}=120$ for VMC) were taken.
The number of mesh points for each angle coordinate was 5,000.
The number of Monte Carlo sampling in the integral calculations of the HF, TC, and BITC methods was up to 2.56 billions for each SCF iteration.
These large numbers might be improved by efficient implementation of the Monte Carlo sampling, while it is not the scope of this study.
An interesting alternative way for grid integration is using Treutler-Ahlrichs integration grids~\cite{grid1, grid2} without Monte Carlo sampling, as discussed in Ref.~\onlinecite{TCatoms_2023}.
The number of SCF cycles was set to 30.
Since it is not easy to judge whether the self-consistency is reached because of the statistical error,
we estimated the total energy as a statistical average for the last 10 SCF loops in the way described later in this section.
Jastrow functions described in Sec.~\ref{sec:jastrow} were used. For a helium atom, only the Jastrow function for the antiparallel spin pairs is required.

For the VMC calculation, we used CASINO code~\cite{CASINO}.
We minimized the unreweighted variance in the Jastrow-parameter optimization~\cite{unreweighted}.
We took at most 500 thousand of Monte Carlo steps and 20 cycles for each optimization.
These relatively large values are required because of the high sensitivity of the TC and BITC results on the Jastrow parameters as we shall see later. We note that the final total energy was evaluated in a separate VMC run using at most 800 million of Monte Carlo steps.

For the TC$+$VMC self-consistent loops, we judged that the self-consistency is reached when the Jastrow parameters are not optimized further. To say, the variance is not lowered in further VMC calculations. Typically 10--20 iterations were required to reach the self-consistency between the TC and VMC calculations.

We show the VMC energy with an error bar estimated by the CASINO code.
For the HF, TC, and BITC methods, we calculated the standard error of the total energies for the last 10 SCF iterations, and show it as an estimated error for the Monte Carlo sampling and the self-consistency within the HF, TC, and BITC calculations.

In the TC$+$VMC method, we can use both the TC and BITC methods in the orbital optimization. However, for a helium atom, we found that the right one-electron orbitals are almost the same between the TC and BITC methods as long as the same Jastrow factor is used. Thus, in this study, we optimized the one-electron orbitals by the TC method until the TC$+$VMC self-consistent loop is converged, and finally evaluated $E_{\mathrm{TC}}$ and $E_{\mathrm{BITC}}$ using the optimized Jastrow function by performing the TC and BITC calculations. Also for other atoms, since we found that right one-electron orbitals are very similar between the TC and BITC methods, we adopt the same way for calculation.
We also found that the imaginary part of the orbitals and eigenvalues are within statistical uncertainty of zero for closed-shell atoms investigated in this study. At the moment, we only numerically verified this and do not have a proof for it, which is an important future issue.

Note that one-electron orbitals used in VMC calculation should be orthonormalized (see Sec.~\ref{sec:orb_orthognal} for theoretical detail of the orthonormalization in the TC and BITC calculation). Thus, we should perform Gram-Schmidt orthonormalization also for the BITC orbitals if one would like to use the BITC orbitals as an input for VMC calculations, while the orthonormalized orbitals are not used in the BITC calculation itself.
This kind of orthogonalization is validated by the invariance of the Slater determinant against a linear combination within the occupied orbitals.

\subsection{Convergence with respect to the number of basis functions}

We first checked convergence with respect to the number of basis functions, $N_{\mathrm{basis}}$.
In this section, we took a twice larger number of Monte Carlo sampling in the HF and TC calculations than that shown in Sec.~\ref{sec:Computationa_details} (i.e., that used in other sections), to reduce the statistical error.

\begin{figure}
\begin{center}
\includegraphics[width=8.0 cm]{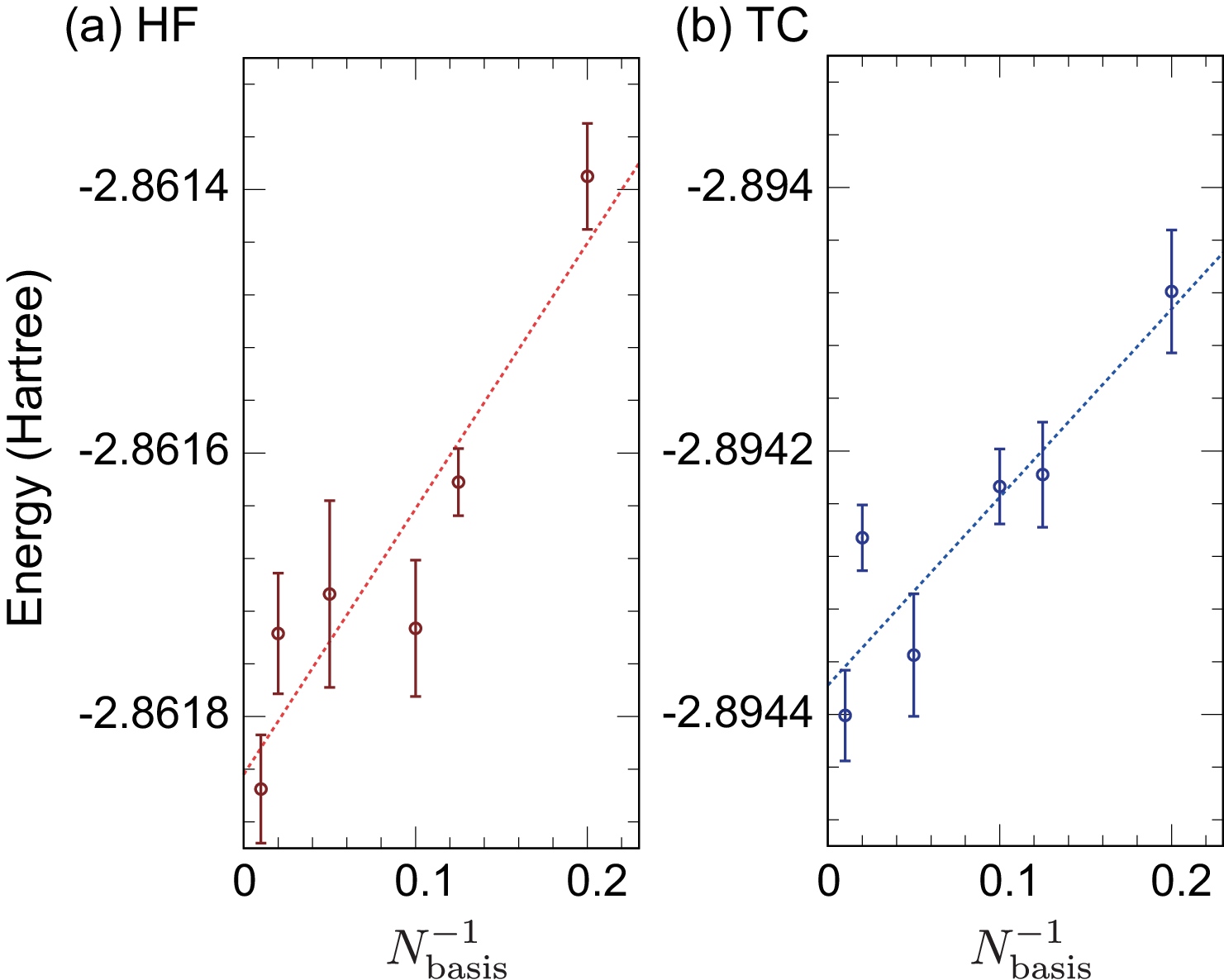}
\caption{Total energy of a helium atom plotted against the inverse of the number of basis functions, $N_{\mathrm{basis}}^{-1}$, (a) for the HF method and (b) for the TC method ($E_{\mathrm{TC}}$) using the $S=S_{\mathrm{minimal}}=\{(1,0,0)\}$ Jastrow function in Eq.~(\ref{eq:Jastrow_pade}) with $a=1.5$ Bohr. Lines are guides for eyes.}
\label{fig:basis}
\end{center}
\end{figure}

Figure~\ref{fig:basis} presents the total energy of a helium atom plotted against the inverse of the number of basis functions, $N_{\mathrm{basis}}^{-1}$, (a) for the HF method and (b) for the TC method ($E_{\mathrm{TC}}$) using the $S=S_{\mathrm{minimal}}=\{(1,0,0)\}$ Jastrow function in Eq.~(\ref{eq:Jastrow_pade}) with $a=1.5$ Bohr.
Based on these plots, we determined to set $N_{\mathrm{basis}}=50$ from the next section both for the HF, TC, and BITC calculations, by which an expected basis-set error is sufficiently small ($\sim 0.1$ mHt.).
We note that our Hartree-Fock energy of a helium atom with $N_{\mathrm{basis}}^{-1}\to0$ (i.e. $N_{\mathrm{basis}}\to\infty$) is consistent with the complete-basis-set limit reported in an old literature ($-2.861679995612$ Ht.)~\cite{HFCBS}. A small difference mainly comes from the statistical error in Monte Carlo sampling and possibly also from the number of spatial mesh points in our calculation, while it is not the scope of our study to see the convergence at the level of $0.1$ mHt.
We do not show the BITC result since it shows the same trend as the TC result.
We verified that $N_{\mathrm{basis}}=50$ is sufficiently large, i.e., the error is $\sim$ 0.1 mHt., also for beryllium and neon atoms.

\subsection{Total energy and orbital energy (He)}

\subsubsection{Total energy~\label{sec:tote}}

Before presenting our calculation results, here we define a notation of the total energies estimated by several different ways, as summarized in Table~\ref{tab:notation}.
For HF$+$VMC calculation, we evaluated the total energy, Eq.~(\ref{eq:totE}), by VMC, which is denoted as $E_{\mathrm{VMC}}^{\mathrm{HF+VMC}}$.
On the other hand, for TC$+$VMC calculation, we have three ways to estimate the total energy. The total energy, Eq.~(\ref{eq:totE}), evaluated by VMC is denoted as $E_{\mathrm{VMC}}^{\mathrm{TC+VMC}}$, and those evaluated by Eqs.~(\ref{eq:EtotTC}) and (\ref{eq:EtotBITC}) are denoted as $E_{\mathrm{TC}}^{\mathrm{TC+VMC}}$ and $E_{\mathrm{BITC}}^{\mathrm{TC+VMC}}$, respectively.
These different estimates of the total energy coincide when the many-body wave function is the exact eigenstate.

\begin{table}
\caption{Notation of the total energies estimated by several different ways. As noted in the main text, for simplicity, we optimized one-electron orbitals by the TC method even for calculating $E_{\mathrm{BITC}}^{\mathrm{TC+VMC}}$ in this study. For calculating $E_{\mathrm{BITC}}^{\mathrm{TC+VMC}}$, BITC calculation was performed only once after TC$+$VMC optimization finished.}
\vspace*{2mm}
\label{tab:notation}
\begin{tabular}{c c c c}
\hline
 & Eq. of the total energy\  & \ orbital opt.\  & \ Jastrow opt.\\
 \hline \hline
 $E_{\mathrm{VMC}}^{\mathrm{HF+VMC}}$ & Eq.~(\ref{eq:totE}) \ ($E_{\mathrm{VMC}}$) & HF & VMC\\
 $E_{\mathrm{VMC}}^{\mathrm{TC+VMC}}$ & Eq.~(\ref{eq:totE}) \ ($E_{\mathrm{VMC}}$) & TC & VMC\\
 $E_{\mathrm{TC}}^{\mathrm{TC+VMC}}$ & Eq.~(\ref{eq:EtotTC}) \ ($E_{\mathrm{TC}}$) & TC & VMC\\
 $E_{\mathrm{BITC}}^{\mathrm{TC+VMC}}$ & Eq.~(\ref{eq:EtotBITC}) \ ($E_{\mathrm{BITC}}$)  & TC & VMC\\
 \hline
 \end{tabular}
 \end{table}

\begin{figure*}
\begin{center}
\includegraphics[width=12 cm]{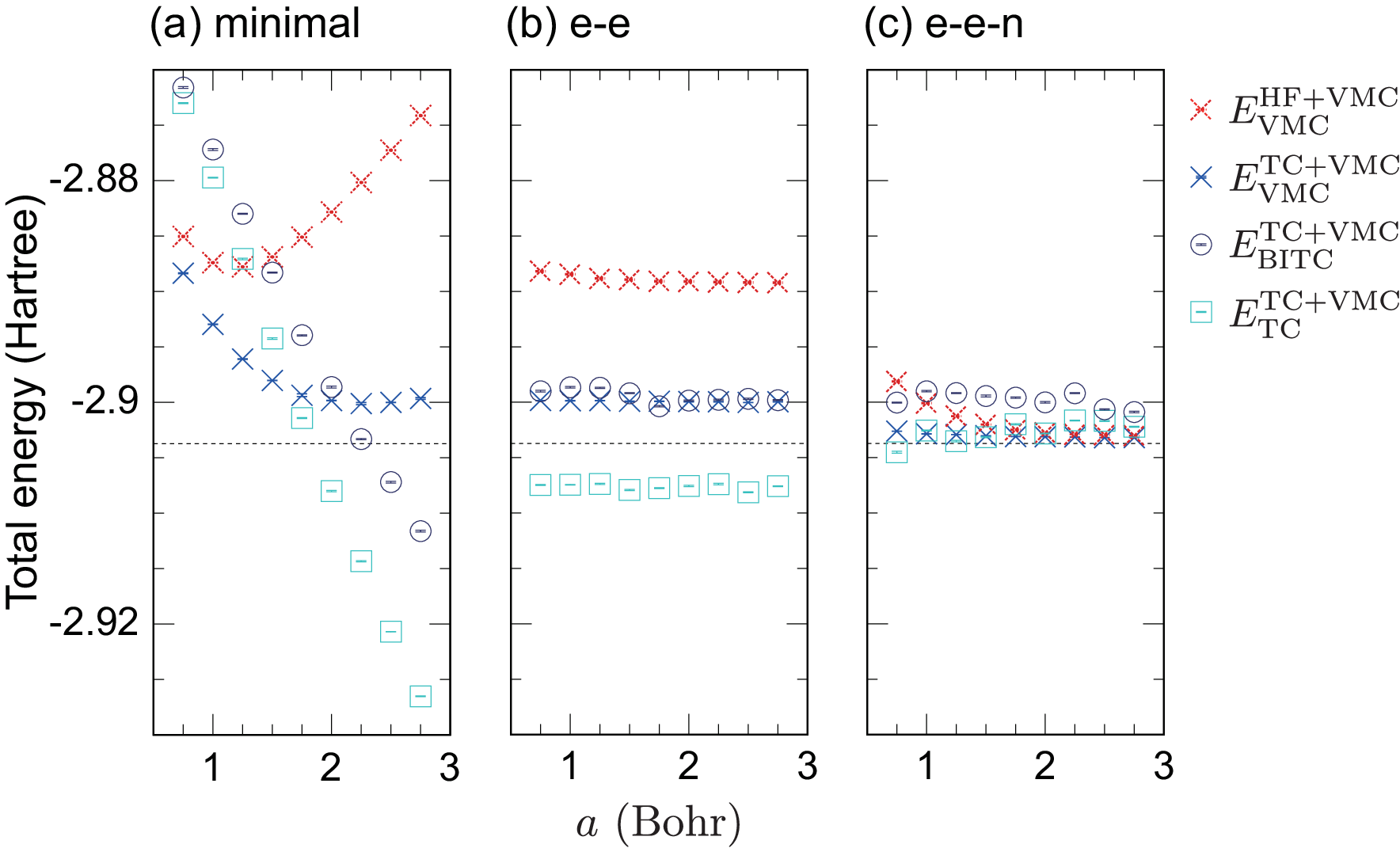}
\caption{Calculated total energies of a helium atom using the Jastrow functions with (a) $S=S_{\mathrm{minimal}}$, (b) $S=S_{\mathrm{ee}}$, (c) $S=S_{\mathrm{een}}$, respectively. The broken lines show the exact total energy ($E_{\mathrm{exact}}=-2.90372$ Ht.) estimated in Ref.~\onlinecite{He_totEex}.}
\label{fig:tote}
\end{center}
\end{figure*}

We compare the calculated total energies described in the previous paragraph for a helium atom in Fig.~\ref{fig:tote}.
These values are also listed in Table~\ref{tab:tote}, where $a=1.5$ Bohr for the $S_{\mathrm{ee}}$ and $S_{\mathrm{een}}$ Jastrow functions are chosen as representative cases.
Figure \ref{fig:tote}(a) shows that the orbital optimization by the TC method successfully improves a quality of many-body wave functions, because $E_{\mathrm{VMC}}^{\mathrm{TC+VMC}}$ is lower than $E_{\mathrm{VMC}}^{\mathrm{HF+VMC}}$.
This improvement is also found for the $S_{\mathrm{ee}}$ Jastrow function as shown in Fig.~\ref{fig:tote}(b). Because higher-order polynomial terms are included for the $S_{\mathrm{ee}}$ Jastrow function, the total energy less depends on the Jastrow parameter $a$ (shown as the horizontal axis) in Fig.~\ref{fig:tote}(b) unlike in Fig.~\ref{fig:tote}(a). It is also noteworthy that the best $a$ parameter in Fig.~\ref{fig:tote}(a) provides a similar $E_{\mathrm{VMC}}$ to that in Fig.~\ref{fig:tote}(b), which might be due to a simplicity of the Jastrow function for a helium atom. When using the $S_{\mathrm{een}}$ Jastrow function, $E_{\mathrm{VMC}}^{\mathrm{TC+VMC}}$ and $E_{\mathrm{VMC}}^{\mathrm{HF+VMC}}$ are very similar as shown in Fig.~\ref{fig:tote}(c). This is not necessarily the case for other atoms where nodal one-body wave functions are included in the Slater determinant, because a nodal structure cannot be represented with the Jastrow factor. The case of other closed-shell atoms shall be discussed in Sec.~\ref{sec:other_atoms}.
In Table~\ref{tab:tote}, the total energy obtained using one-body and electron-electron Jastrow terms reported in Ref.~\cite{He_VMC} is similar to that obtained in our TC$+$VMC calculation using $S_{\mathrm{ee}}$ Jastrow. This is naturally understood because the effect of orbital relaxation can be included through the one-body Jastrow terms for He where one-electron orbitals are nodeless. The total energy obtained using one-body, electron-electron, and electron-electron-nucleus Jastrow terms reported in Ref.~\cite{He_VMC} is lower than that obtained in our TC$+$VMC calculation using $S_{\mathrm{een}}$ Jastrow because our calculation used fewer number of variational parameters in the Jastrow factor.

While $E_{\mathrm{VMC}}$ is variational, $E_{\mathrm{TC}}$ and $E_{\mathrm{BITC}}$ are not so because of the non-Hermiticity of the TC Hamiltonian. Related to this feature, although evaluated for the same many-body wave function, $E_{\mathrm{VMC}}^{\mathrm{TC+VMC}}$, $E_{\mathrm{TC}}^{\mathrm{TC+VMC}}$, and $E_{\mathrm{BITC}}^{\mathrm{TC+VMC}}$ are much different in Fig.~\ref{fig:tote}(a).
This discrepancy becomes smaller by improving the Jastrow factor, as shown in Fig.~\ref{fig:tote}(b)(c).
In total, $E_{\mathrm{TC}}^{\mathrm{TC+VMC}}$ tends to be too low, i.e. overcorrelated, for low-quality Jastrow factors, but it is somewhat alleviated for  $E_{\mathrm{BITC}}^{\mathrm{TC+VMC}}$ (also see Table~\ref{tab:tote}). It is likely because the left (bra) orbital in the BITC method tends to delocalize compared with the HF orbital, which is the opposite trend to the right (ket) orbital in the TC and BITC method, as we shall see in Sec.~\ref{sec:orbital_shape}.
The TC total energy is overcorrelated also for solid-state calculation~\cite{TCMP2}.

One important problem we found is a high sensitivity of $E_{\mathrm{TC}}^{\mathrm{TC+VMC}}$ and $E_{\mathrm{BITC}}^{\mathrm{TC+VMC}}$ to the Jastrow parameters.
This is seen, e.g., from a large variation of $E_{\mathrm{TC}}^{\mathrm{TC+VMC}}$ and $E_{\mathrm{BITC}}^{\mathrm{TC+VMC}}$ for $2\leq a \leq 3$ Bohr in Fig.~\ref{fig:tote}(a), while $E_{\mathrm{VMC}}^{\mathrm{TC+VMC}}$ evaluated using the same right one-electron orbitals is almost unchanged against the $a$ parameter. We speculate that such high sensitivity is to some extent relevant to the non-Hermiticity of $\mathcal{H}_{\mathrm{TC}}$ discussed above that breaks the variational principle and then the variation of $E_{\mathrm{TC}}^{\mathrm{TC+VMC}}$ and $E_{\mathrm{BITC}}^{\mathrm{TC+VMC}}$ is not restricted by the lower bound, i.e., the exact total energy.
The energy curves of $E_{\mathrm{TC}}^{\mathrm{TC+VMC}}$ and $E_{\mathrm{BITC}}^{\mathrm{TC+VMC}}$ in Fig.~\ref{fig:tote}(c) are not very smooth against the $a$ parameter, because of the high-sensitivity of those energies to the Jastrow parameters $c_{ijk}^{\sigma_1\sigma_2}$, which makes a difficulty in achieving the self-consistency for the TC$+$VMC calculations. 
This is the reason for not optimizing the $a$ parameter in VMC calculation. By taking the $a$ parameter as the horizontal axis in Fig.~\ref{fig:tote}, we can check whether the self-consistency between the TC and VMC optimizations was successfully achieved and the solution did not fall into peculiar local minima. If something goes wrong in achieving self-consistency, the energy will exhibit an abrupt change against the $a$ parameter.
One possible way to evade this instability is non-self-consistent approach where the Jastrow parameters optimized for the HF orbitals are used in the TC calculation, as adopted in Ref.~\cite{TCatoms_HFJastrow}. We shall discuss the effect of the iterative TC and VMC calculations (to say, the self-consistency in the TC$+$VMC calculation) in Sec.~\ref{sec:oneshot}.

We here make a few comments on the consistency with the previous study by Umezawa $et$ $al$.~\cite{Umezawa}
They calculated the total energies, $E_{\mathrm{VMC}}^{\mathrm{HF+VMC}}$, $E_{\mathrm{VMC}}^{\mathrm{TC+VMC}}$. and $E_{\mathrm{TC}}^{\mathrm{TC+VMC}}$, by using $S_{\mathrm{minimal}}$ Jastrow function with the best $a$ parameter ($a=1.92$ Bohr).
The ratio of the retrieved correlation energies were reported as 61\%, 90\%, and 77\% for $E_{\mathrm{VMC}}^{\mathrm{HF+VMC}}$, $E_{\mathrm{VMC}}^{\mathrm{TC+VMC}}$, and $E_{\mathrm{TC}}^{\mathrm{TC+VMC}}$, respectively.
We calculated these values using the same Jastrow function, and obtained 53.5\%, 90.3\%, and 106.8\%, respectively.
They are roughly consistent except for $E_{\mathrm{TC}}^{\mathrm{TC+VMC}}$.
One possible reason for the difference of $E_{\mathrm{TC}}^{\mathrm{TC+VMC}}$ is a basis-set error for the one-electron orbitals in the previous study.
Because of the coincidence of $E_{\mathrm{VMC}}$ and $E_{\mathrm{TC}}$ in the $S_{\mathrm{een}}$ Jastrow function as shown in Fig.~\ref{fig:tote}(c), we consider that $E_{\mathrm{TC}}^{\mathrm{TC+VMC}}$ calculated by us using a larger number of basis functions is reliable.
Another important difference between Umezawa $et$ $al$. and ours is that the fact that $\sigma^{2}_{\mathrm{TC}}$ is minimized in their calculation, as mentioned in Sec.~\ref{sec:TCVMC}. Regarding this point, the best $a$ parameter for $S_{\mathrm{minimal}}$ in Ref.~\cite{Umezawa}, $a=1.92$ Bohr, seems to be consistent with our calculation (see Fig.~\ref{fig:tote}(a)), which suggests that $\sigma^{2}_{\mathrm{TC}}$ minimization also works well as a guiding principle for Jastrow optimization.
In fact, it was recently reported that $\sigma^{2}_{\mathrm{TC}}$ minimization in VMC offers accurate and stable optimization of the Jastrow parameters~\cite{TCatoms_2023}.

\begin{table*}
\caption{Total energy (Ht.) and the ratio of the correlation energy retrieved (\%) for a helium atom. For evaluating the latter quantity, the HF energy calculated by us using $N_{\mathrm{basis}}=50$ ($E_{\mathrm{HF}}=-2.8617(3)$ Ht.) and the exact total energy ($E_{\mathrm{exact}}=-2.90372$ Ht.) estimated in Ref.~\onlinecite{He_totEex} were used. For all data except those taken from Ref.~\cite{Umezawa}, $a=1.5$ Bohr was used in the Jastrow factor. For data taken from Ref.~\cite{Umezawa}, $a=1.92$ Bohr was used as the best value obtained by variance ($\sigma^2_{\mathrm{TC}}$) minimization. The statistical error for our calculation results is not shown in this table because it is much less than the expected basis-set error $\sim 0.1$ mHt. We note that the variational principle breaks for $E_{\mathrm{TC}}$ and $E_{\mathrm{BITC}}$, which can result in the ratio of the correlation energy retrieved less than 0 \% or larger than 100 \%. For VMC calculations in Ref.~\cite{He_VMC}, a detail of variational parameters is shown in that reference.}
\vspace*{2mm}
\label{tab:tote}
\begin{tabular}{c c c c c c c c c}
\hline
Method & Jastrow& $E_{\mathrm{VMC}}$ & (\%) & $E_{\mathrm{TC}}$ & (\%) & $E_{\mathrm{BITC}}$ & (\%) & Ref. \\
 \hline \hline
HF$+$VMC & $S_{\mathrm{ee}}$ & $-2.8889$ & $64.8$ & - & - & - & - & -\\
  & $S_{\mathrm{een}}$ & $-2.9020$ & $95.9$ & - & - & - & - & -\\
TC$+$VMC & $S_{\mathrm{ee}}$ & $-2.8999$ & $90.9$ & $-2.9079$ & $110.0$ & $-2.8992$ & $89.2$& -\\
  & $S_{\mathrm{een}}$ & $-2.9030$ & $98.4$ & $-2.9032$ & $98.7$ & $-2.8994$ & $89.8$& -\\
 \hline
HF$+$VMC & $S_{\mathrm{minimal}}$ & $-2.8873(8)$ & $61$ & - & - & - & - & \cite{Umezawa}\\
 TC$+$VMC & $S_{\mathrm{minimal}}$ & $-2.8997(4)$ & $90$ &  $-2.8942$ & $77$ & - & - & \cite{Umezawa}\\
 \multicolumn{2}{c}{VMC (one-body and e-e Jastrow)} & $-2.900010(9)$ & $91.17(2)$ & - & - & - & - & \cite{He_VMC} \\
  \multicolumn{2}{c}{VMC (one-body, e-e, and e-e-n Jastrow)} & $-2.903693(1)$ & $99.926(2)$ & - & - & - & - & \cite{He_VMC} \\
 \hline
 \end{tabular}
 \end{table*}

\subsubsection{Orbital energy}

Because the TC and BITC method can be regarded as a single-Slater-determinant (HF) approximation for the similarity-transformed Hamiltonian,
we can naturally obtain the orbital energies in Eq.~(\ref{eq:SCF}), and Re[$\epsilon_{ii}$] satisfies the Koopmans' theorem as proved in Ref.~\cite{Umezawa}.
In this section, we investigate the accuracy of the orbital energy of a helium atom.
From this section, we consider the $S_{\mathrm{ee}}$ and $S_{\mathrm{een}}$ Jastrow functions using $a=1.5$ Bohr as representative cases, because of a small $a$-dependence of the total energy in Fig.~\ref{fig:tote}(b)(c).

Table~\ref{tab:orbe} presents the calculated ionization potential (IP) estimated from the $1s$ orbital energy for a helium atom.
We can see that IP estimated from HF is the closest to the exact value, and TC and BITC tend to overestimate it.
The reason can be understood by the following reason. It is well known that, in the HF method, two errors of IP are partially canceled out: one is a lack of the correlation effects and the other one is a lack of the orbital relaxation that actually takes place when an electron is removed from an atom.
By the lack of the correlation effects, the HF total energy of He is overestimated while the total energy of the single-electron system He$^+$ is not, which results in underestimation of IP ($=E(\mathrm{He}^+)-E(\mathrm{He})$).
On the other hand, by the lack of the orbital relaxation, the total energy of He$^+$ is overestimated, which results in overestimation of IP. These two errors are partially canceled in IP evaluated with the Hartree-Fock orbital energy $\epsilon_{ii}$ of He.
However, as for TC and BITC, an error of the correlation effects will be smaller while an error of the orbital relaxation will still take place, resulting in overestimation of IP.
Note that this is not the failure of the TC and BITC methods themselves, but rather due to the limitation of the Koopmans' theorem.
The ionization potentials of 0.9180 Ht. for HF and 0.9179 Ht. for TC using $a=1.92$ Bohr for $S_{\mathrm{minimal}}$ were reported in Ref.~\cite{Umezawa}, the former of which is consistent with ours while we obtained 0.9290 Ht. for TC using the same Jastrow factor.
A discrepancy of IP for TC is likely relevant to the difference in $E_{\mathrm{TC}}$ as discussed in Sec.~\ref{sec:tote}, the origin of which might be the basis-set error in the previous study, Ref.~\cite{Umezawa}. 

\begin{table}
\caption{Ionization potential (IP) (Ht.) estimated from the $1s$ orbital energy for a helium atom. The exact IP is determined from a difference between the exact total energy of helium atom, $E_{\mathrm{exact}}=-2.90372$ Ht., estimated in Ref.~\onlinecite{He_totEex}, and the exact total energy of He$^{+}$, $-2.0$ Ht. For all data except those taken from Ref.~\cite{Umezawa}, $a=1.5$ Bohr was used in the Jastrow factor. For data taken from Ref.~\cite{Umezawa}, $a=1.92$ Bohr was used as the best value obtained by variance ($\sigma^2_{\mathrm{TC}}$) minimization. The statistical error for our calculation results is not shown in this table because it is much less than the expected basis-set error $\sim 0.1$ mHt.}
\vspace*{2mm}
\label{tab:orbe}
\begin{tabular}{c c c c}
\hline
Method & Jastrow & IP & Ref.\\
 \hline \hline
HF &- & $0.9180$ & -\\
 & - & $ 0.9180$ & \cite{Umezawa}\\
\hline
TC & $S_{\mathrm{ee}}$ & $0.9301$& -\\
 & $S_{\mathrm{een}}$ & $0.9352$& -\\
BITC & $S_{\mathrm{ee}}$ & $0.9507$& -\\
  & $S_{\mathrm{een}}$ & $0.9531$& -\\
 \hline
TC &  $S_{\mathrm{minimal}}$ & $0.9179$ & \cite{Umezawa}\\
 \hline
Exact & - & $0.90372$ & \cite{He_totEex}\\
 \hline
 \end{tabular}
 \end{table}

\subsection{Optimized many-body wave functions (He)}
\subsubsection{Optimized Jastrow functions\label{sec:opt_jastr_funct}}

\begin{figure}
\begin{center}
\includegraphics[width=8.4 cm]{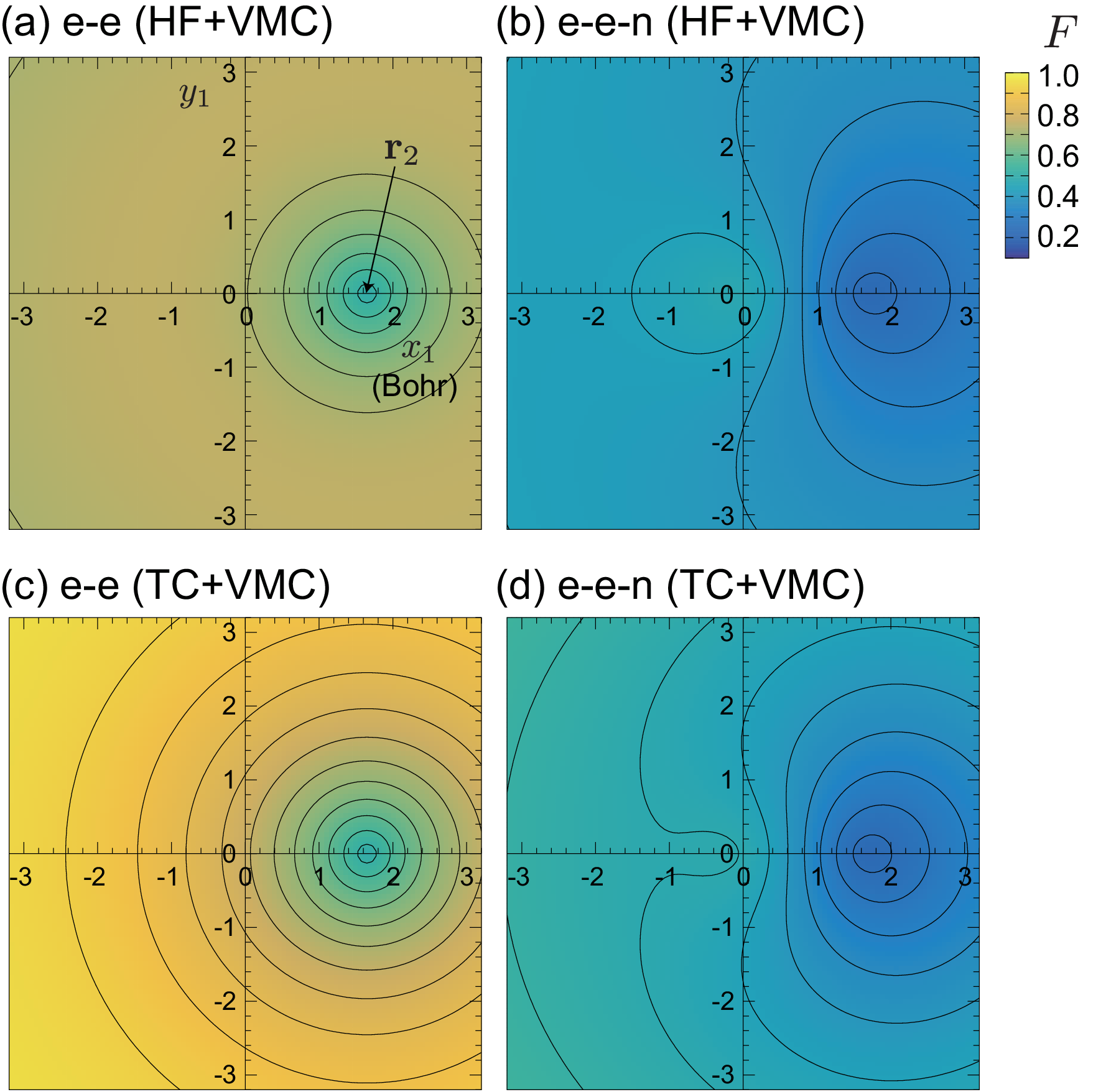}
\caption{Optimized Jastrow factor $F(x_1,x_2)=\mathrm{exp}(-u(x_1,x_2))$, of a helium atom with constraints of $z_1=0$, $\mathbf{r}_2=(1.65, 0, 0)$ (Bohr), and $\sigma_1 = -\sigma_2$ (see the main text). The $S_{\mathrm{ee}}$ and $S_{\mathrm{een}}$ Jastrow functions were used for panels (a)(c) and (b)(d), respectively. The Jastrow factors optimized by HF$+$VMC and TC$+$VMC are shown in panels (a)(b) and (c)(d), respectively. For all the panels, $a=1.5$ Bohr was used.}
\label{fig:optjas}
\end{center}
\end{figure}

Figure~\ref{fig:optjas} presents the optimized Jastrow factor, $F(x_1,x_2)=\mathrm{exp}(-u(x_1,x_2))$, of a helium atom for several conditions.
The position of one electron, $\mathbf{r}_2$, is fixed at $\mathbf{r}_2=(1.65, 0, 0)$ (Bohr) as shown in Fig.~\ref{fig:optjas}(a).
In addition, the $z$ coordinate of $\mathbf{r}_1=(x_1,y_1,z_1)$ is fixed at $0$, i.e., $z_1=0$, in this plot.
The spin coordinates of two electrons are assumed to be antiparallel ($\sigma_1 = -\sigma_2$).

\begin{figure}
\begin{center}
\includegraphics[width=8.4 cm]{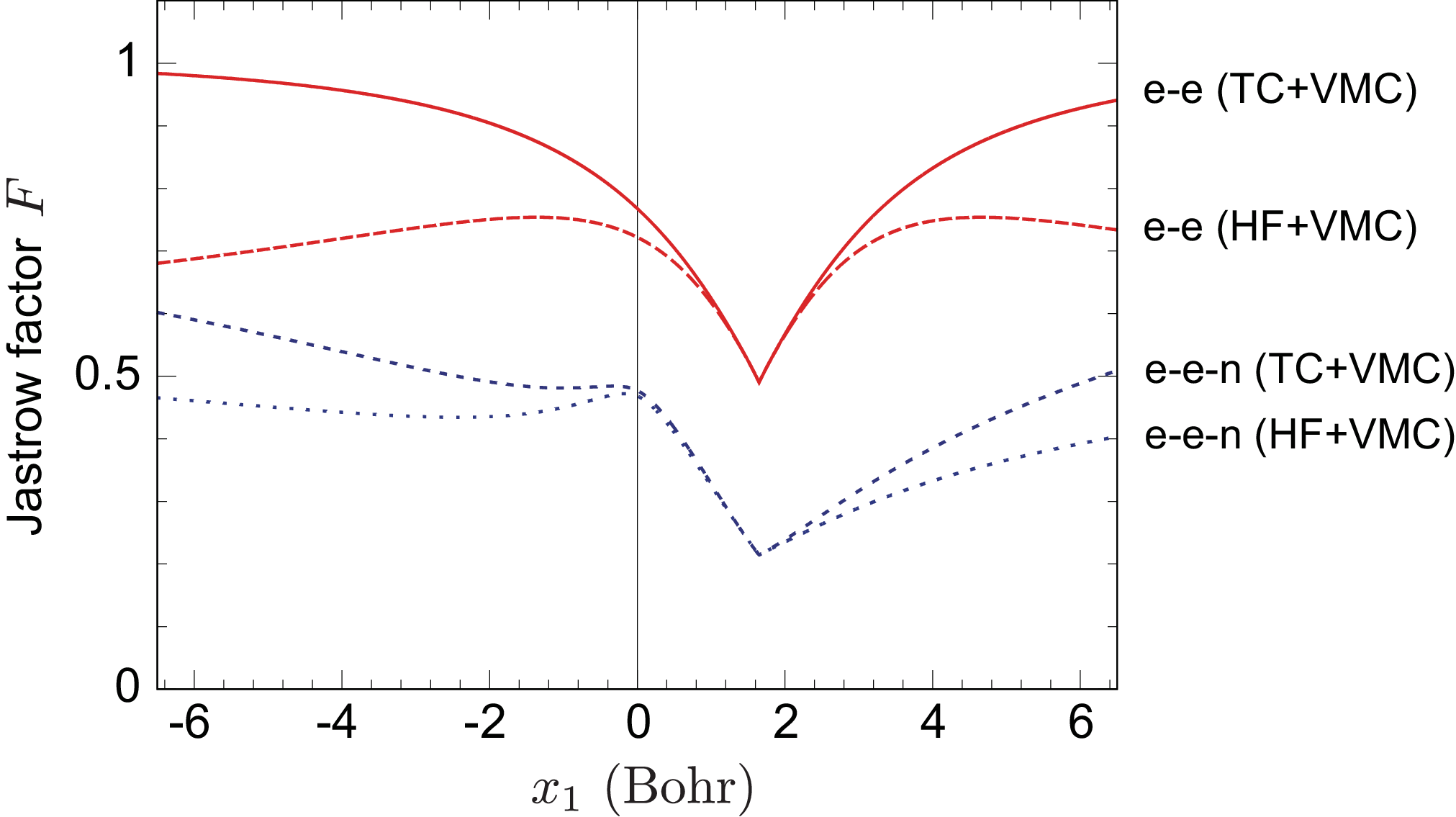}
\caption{The same plot as Fig.~\ref{fig:optjas} on the $y_1=z_1=0$ line. Each line represents the Jastrow factor $F$ shown in Figs.~\ref{fig:optjas}(a)--(d).}
\label{fig:optjas_line}
\end{center}
\end{figure}

In Figs.~\ref{fig:optjas}(a)(c), contour lines are concentric circles because the Jastrow factor only has a $|\mathbf{r}_1-\mathbf{r}_2|$ dependence. 
A small weight of the Jastrow factor near $\mathbf{r}_1 \sim \mathbf{r}_2$ represents that electrons avoid each other by strong Coulomb repulsion. By including the $S_{\mathrm{een}}$ Jastrow terms, in Figs.~\ref{fig:optjas}(b)(d), the Jastrow function can exhibit more complex behavior.
Here, a weight of the Jastrow factor near the nucleus increases to some extent compared with Figs.~\ref{fig:optjas}(a)(c), which is a natural consequence of the electron-nucleus attractive interaction. In other words, the electron-electron-nucleus terms in the Jastrow function alleviate the overscreening caused by the $S_{\mathrm{ee}}$ Jastrow function, which does not take into account the position of the nucleus.  We can see this trend also in Fig.~\ref{fig:optjas_line}, where the same plots constrained on the $y_1=z_1=0$ line are shown.

Note that, in Figs.~\ref{fig:optjas} and \ref{fig:optjas_line}, a constant multiplication of the Jastrow factor $F$ is physically meaningless because this degree of freedom is absorbed in the normalization condition of the many-body wave function. Even though, the VMC total energy, $\langle \Phi F| \mathcal{H} | F\Phi\rangle/\langle F\Phi | F\Phi \rangle$ is appropriately defined. Also in the TC and BITC methods, not $F$ but $F^{-1}\mathcal{H}F$ appears in equations, where such a constant multiplication does not play any role.
Therefore, in Figs.~\ref{fig:optjas} and \ref{fig:optjas_line}, we multiply a constant for the Jastrow factor $F$ obtained by HF$+$VMC so that it has almost the same value at the electron-electron coalescence point as that obtained by TC$+$VMC using the same type of the Jastrow parameters (e-e or e-e-n). For example, in Fig.~\ref{fig:optjas_line}, Jastrow factors of e-e (HF$+$VMC) and e-e (TC$+$VMC) almost coincide at the electron-electron coalescence point.

By comparing the Jastrow factors obtained by HF$+$VMC and TC$+$VMC, we can see that the Jastrow factors obtained by TC$+$VMC tend to have a larger weight in the region where an electron-electron distance is long.
On the other hand, TC one-electron orbitals tend to be more localized than HF orbitals as we shall see later.
Thus, it seems that the Jastrow factor and one-electron orbitals (Slater determinant) exhibit an inverse trend to balance it out for well-approximating the exact eigenstate.

\begin{figure*}
\begin{center}
\includegraphics[width=15.5 cm]{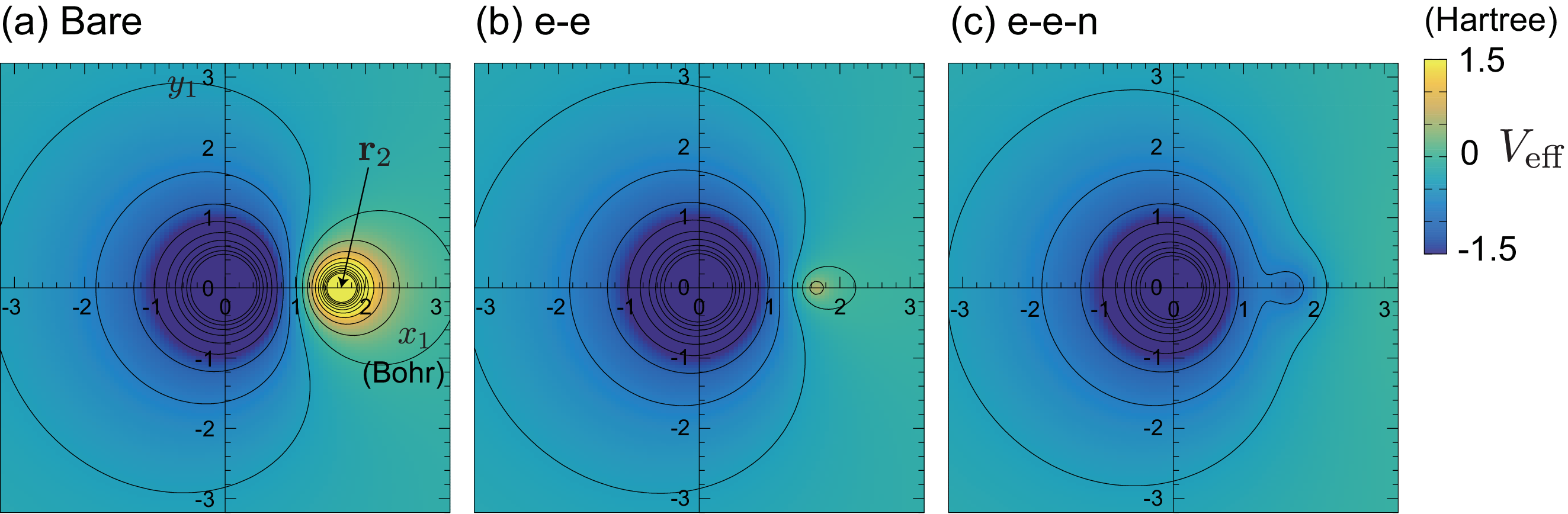}
\caption{TC effective potential $V_{\mathrm{eff}}(x_1,x_2)$ as defined in Eq.~(\ref{eq:Veff}), with constraints of $z_1=0$, $\mathbf{r}_2=(1.65, 0, 0)$ (Bohr), and $\sigma_1 = -\sigma_2$ (see the main text), for a helium atom. (a) The bare potential, to say, $V_{\mathrm{eff}}$ without $\nabla^2 u$ terms, is shown, instead of the TC effective potential. (b)(c) The TC effective potential for several conditions, which are the same as those for Fig.~\ref{fig:optjas}(c)(d). Contour lines are shown as guides for eyes.}
\label{fig:effpot}
\end{center}
\end{figure*}

To see the role of the Jastrow factor in more detail, in Fig.~\ref{fig:effpot}, we show the TC effective potential of the electron at $x_1=(\mathbf{r}_1, \sigma_1)$ defined as follows:
\begin{gather}
V_{\mathrm{eff}}(x_1,x_2) \equiv \notag\\
-\frac{2}{|\mathbf{r}_1|} + \frac{1}{|\mathbf{r}_1-\mathbf{r}_2|}+\frac{\nabla_1^2 u(x_1,x_2)+\nabla_2^2 u(x_1,x_2)}{2}, \label{eq:Veff}
\end{gather}
where the first and second terms represent the electron-nucleus and electron-electron Coulomb potentials, respectively. Here we took a part of the TC effective interaction terms that describes the electron-electron cusp condition, to say, that exhibits a divergent behavior near the electron-electron coalescence point. Other effective potential terms have a relatively minor role near the electron-electron coalescence point since they do not exhibit such a divergent behavior, and are not shown here for simplicity.
Details for plotting, such as the fixed $\mathbf{r}_2$, are the same as those used for Fig.~\ref{fig:optjas}.

\begin{figure}
\begin{center}
\includegraphics[width=8.0 cm]{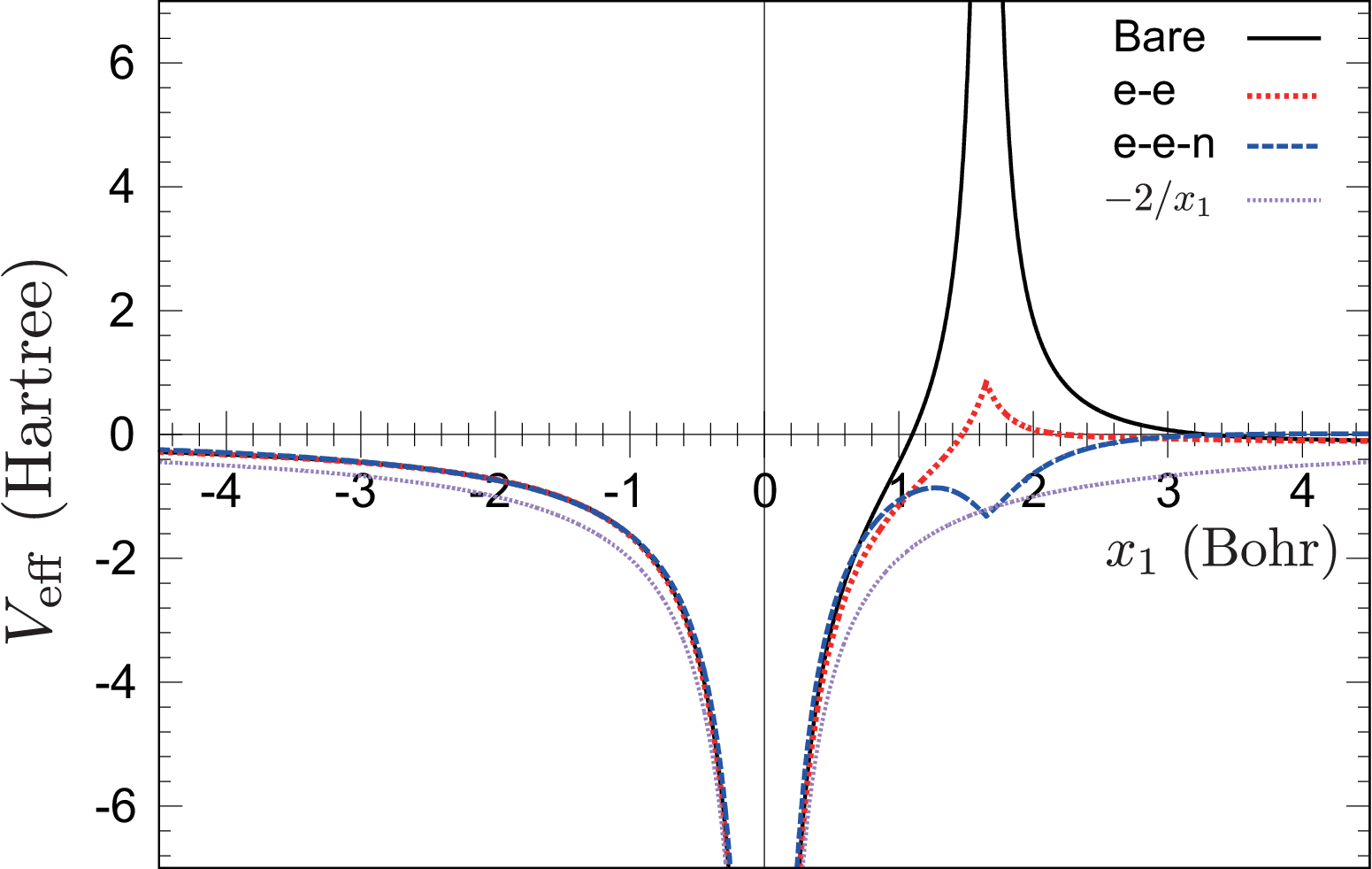}
\caption{The same plot as Fig.~\ref{fig:effpot} on the $y_1=z_1=0$ line, corresponding to Figs.~\ref{fig:effpot}(b)(c).
The Coulomb potential from the nucleus, $-2/x_1$, is shown as a guide for eyes.}
\label{fig:effpot_line}
\end{center}
\end{figure}

Comparing with Fig.~\ref{fig:effpot}(a), where the `bare' potential (i.e. $V_{\mathrm{eff}}$ without $\nabla^2 u$ terms) is shown, we can clearly see, in Figs.~\ref{fig:effpot}(b)(c), that the TC effective potential terms $\nabla^2 u$ successfully cancel out the divergence of the bare electron-electron Coulomb interaction at the electron-electron coalescence point.
Under the weak effective electron-electron interaction, the mean-field (HF) approximation of the TC Hamiltonian is expected to work well. This is the way how the TC method takes into account the electron correlation effects.
The same plot constrained on the $y_1=z_1=0$ line is shown in Fig.~\ref{fig:effpot_line}.

In addition to the above-mentioned trend, we can also see that the effective potential for the $S_{\mathrm{een}}$ Jastrow function is a bit larger than that for the $S_{\mathrm{ee}}$ Jastrow function, and is rather close to the bare potential, near the nucleus (e.g. $|x_1|\leq 1$ Bohr).
This means that the $S_{\mathrm{een}}$ Jastrow terms alleviate the overscreening near the nucleus by the $S_{\mathrm{ee}}$ Jastrow function, as we have seen in Figs.~\ref{fig:optjas} and \ref{fig:optjas_line}.

\subsubsection{Optimized one-electron orbitals\label{sec:orbital_shape}}

\begin{figure}
\begin{center}
\includegraphics[width=8.2 cm]{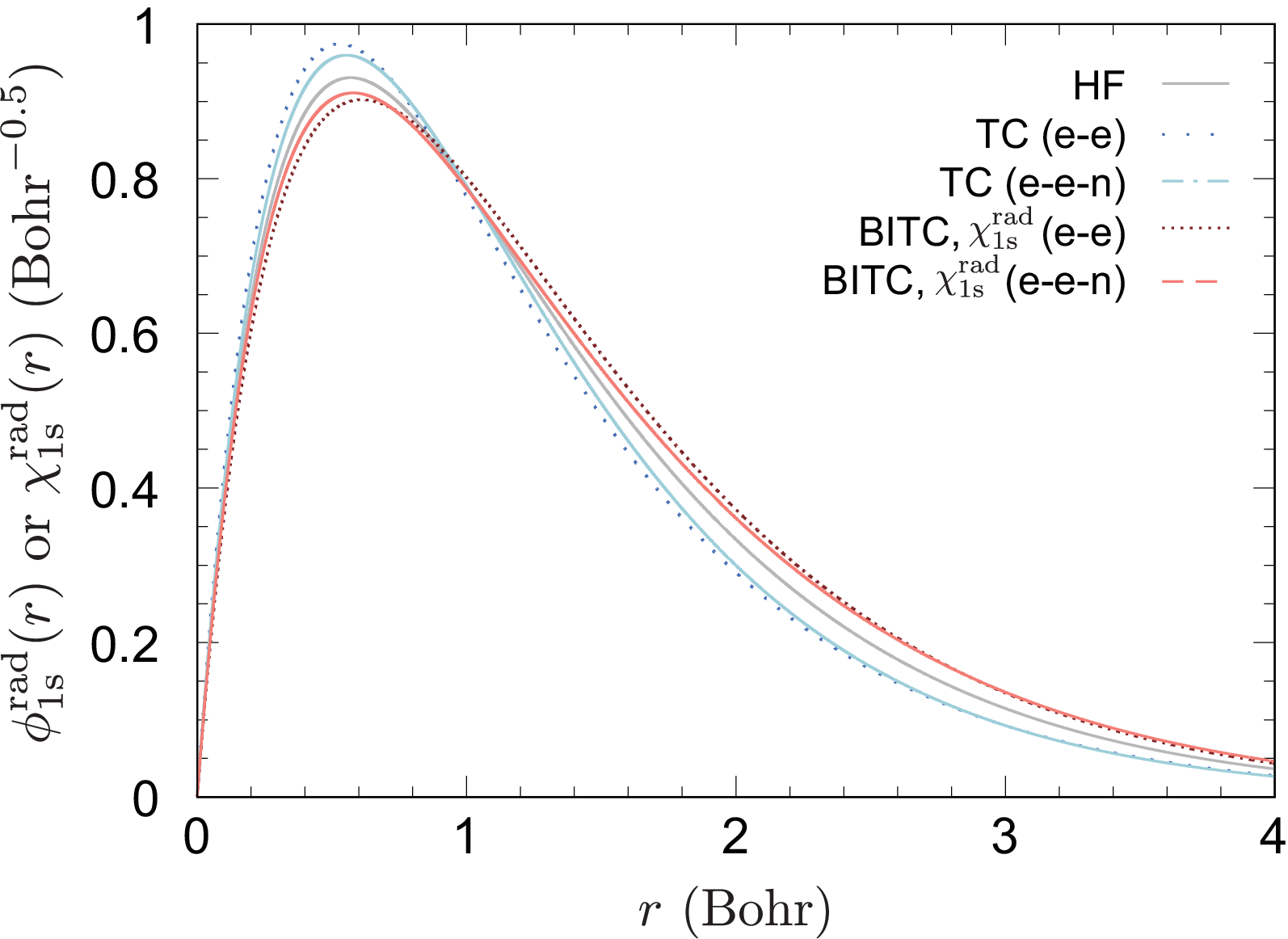}
\caption{The radial wave function for the $1s$ state of a helium atom.The left orbital $\chi$ is shown for the BITC method. The right orbital $\phi$ for the BITC method are not shown since a difference of $\phi$ between TC and BITC is almost discernible.}
\label{fig:optorb}
\end{center}
\end{figure}

Figure~\ref{fig:optorb} presents the optimized one-electron orbitals of the $1s$ state.
Here, the left orbital $\chi$ is shown for the BITC method because a difference of $\phi$ between TC and BITC is almost discernible.
We can clearly see that the one-electron orbital is deformed by optimization in the TC and BITC methods.
Overall, the right orbital $\phi$ is a bit localized compared with the HF orbital, while the left orbital $\chi$ is rather a bit delocalized.
The reason why the right orbital $\phi$ is localized in the TC and BITC methods is as follows: the electron-electron interaction is screened by the Jastrow factor and then an electron is allowed to become closer to the nucleus to get stabilized.
It seems that the one-electron orbital is a bit over-localized for the $S_{\mathrm{ee}}$ Jastrow function, while it is somewhat weakened for the $S_{\mathrm{een}}$ Jastrow function, as is consistent with our observation discussed in Sec.~\ref{sec:opt_jastr_funct}.
The opposite trend for localization between $\phi$ and $\chi$ originates from the fact that the left orbital can be regarded as the mean-field solution of $F\mathcal{H}F^{-1}$ rather than $\mathcal{H}_{\mathrm{TC}} = F^{-1}\mathcal{H}F$, where the effect of the Jastrow factor is expected to be inverted.

\subsection{Effect of the self-consistency between the Jastrow- and orbital-optimizations (He)\label{sec:oneshot}}
We have investigated the self-consistent TC$+$VMC method so far in this paper, where Jastrow parameters and one-electron orbitals are alternately optimized by VMC and TC calculations, respectively, as depicted in Fig.~\ref{fig:scheme}(a).
On the other hand, one-shot TC$+$VMC calculation in the following procedure is also possible:
(i) perform the HF calculation to get the HF orbitals, (ii) perform the VMC calculation to get the optimized Jastrow factor using the HF orbitals, (iii) perform the TC method using the Jastrow parameters obtained in (ii), and (iv) evaluate $E_{\mathrm{VMC}}$ for the Jastrow parameters obtained in (ii) and the TC orbitals obtained in (iii). This procedure is shown in Fig.~\ref{fig:scheme}(b). 
For the one-shot TC$+$VMC calculation, we did not perform an alternate repetition of VMC and TC calculations as was done in the self-consistent TC$+$VMC calculations shown in the previous sections. In other words, the Jastrow parameters optimized for the HF orbitals were used in the one-shot TC$+$VMC calculations, as adopted in Ref.~\cite{TCatoms_HFJastrow}. Note that the self-consistency for solving the one-body SCF equation, Eq.~(\ref{eq:SCF}), in the HF or TC method is always satisfied.

Table~\ref{tab:oneshot} summarizes the calculation results of one-shot TC$+$VMC compared with self-consistent TC$+$VMC.
We can see that one-shot TC$+$VMC to some extent improves a quality of the many-body wave function, i.e., lowers $E_{\mathrm{VMC}}$, and so can be a good alternative way when one would like to reduce computational cost.
On the other hand, it is also clear that the quality of the many-body wave function for one-shot TC$+$VMC is always inferior to that for self-consistent TC$+$VMC. 
For a very simple system like a helium atom, TC$+$VMC calculation using the $S_{\mathrm{minimal}}$ Jastrow function with the optimal $a$ can also be a good alternative of that using the (self-consistently optimized) $S_{\mathrm{ee}}$ Jastrow function, because the former does not require a self-consistent optimization of the Jastrow parameters but offers the same accuracy as the latter as discussed in Sec.~\ref{sec:tote}.
Note that this strategy works well only for simple systems where the $S_{\mathrm{minimal}}$ Jastrow function well approximates the optimized one.

\begin{table*}
\caption{Total energy (Ht.), the ratio of the correlation energy retrieved (\%), and the ionization potential (IP) (Ht.) estimated from the $1s$ orbital energy for each method, for a helium atom. Details are the same as those for Tables~\ref{tab:tote} and \ref{tab:orbe}. Calculated results for HF$+$VMC and self-consistent TC$+$VMC are taken from Tables~\ref{tab:tote} and \ref{tab:orbe}. IP was evaluated by HF, TC, or BITC calculations (i.e., not by VMC). The statistical error for our calculation results is not shown in this table because it is much less than the expected basis-set error $\sim 0.1$ mHt. As noted in Table~\ref{tab:notation}, BITC calculation was performed only once after TC$+$VMC calculation, i.e., using the Jastrow parameters determined by TC$+$VMC calculation. }
\vspace*{2mm}
\label{tab:oneshot}
\begin{tabular}{c c c c c c}
\hline
Jastrow & Method  &  \ \  $E_{\mathrm{VMC}}$  \ \  & \ \  (\%) \ \ & \ \ \ \ IP\ \ \ \ & \ \ \ \ IP (BITC)\ \ \ \ \\
 \hline \hline
$S_{\mathrm{ee}}$ & HF$+$VMC & $-2.8889$ & $64.8$ & 0.9180 & -\\
& TC$+$VMC (one-shot)  & $-2.8974$ & $85.0$ & 0.9239 & 0.9406\\
 & TC$+$VMC (self-consistent)  & $-2.8999$ & $90.9$ & 0.9301 & 0.9507\\
 \hline
$S_{\mathrm{een}}$ & HF$+$VMC &  $-2.9020$ & $95.9$ & 0.9180 & -\\
& TC$+$VMC (one-shot) &  $-2.9029$ & $98.0$ & 0.9415 & 0.9500\\
& TC$+$VMC (self-consistent) &  $-2.9030$ & $98.4$  & 0.9352 & 0.9531\\
  \hline
 \end{tabular}
 \end{table*}
 
\subsection{Other atoms (Be and Ne)\label{sec:other_atoms}}

In this subsection, we present our calculation results for beryllium and neon atoms.
Table~\ref{tab:tote_bene} presents the calculated total energy for beryllium and neon atoms.
We can see that $E_{\mathrm{VMC}}$ is systematically improved by improving a quality of the Jastrow factor, i.e.,
$E_{\mathrm{VMC}}$ using the $S_{\mathrm{een}}$ Jastrow function is much accurate than that using the $S_{\mathrm{ee}}$ Jastrow function.
However, this is not necessarily the case for $E_{\mathrm{TC}}$ and $E_{\mathrm{BITC}}$.
Since these estimations of the total energy, $E_{\mathrm{VMC}}$, $E_{\mathrm{TC}}$, and $E_{\mathrm{BITC}}$, should coincide for the exact many-body wave function, such a non-systematic behavior of $E_{\mathrm{TC}}$ and $E_{\mathrm{BITC}}$ might originate from insufficient accuracy in these atoms unlike a helium atom.
Orbital optimization by the TC method seems to work well for Be and Ne using the $S_{\mathrm{ee}}$ Jastrow factor, but not so for Ne using the $S_{\mathrm{een}}$ Jastrow factor.
The total-energy estimation by $E_{\mathrm{TC}}$ and $E_{\mathrm{BITC}}$ is also unsuccessful for Ne using the $S_{\mathrm{een}}$ Jastrow factor.
These observation suggests that insufficient degrees of freedom in the Jastrow factor can cause non-systematic accuracy of the TC method, or perhaps result in some convergence problem, e.g., trapped by the local minimum in the TC$+$VMC self-consistent optimization.
This can happen because TC and VMC calculations are based on different guiding principles for optimizing many-body wave functions.
For Ne, it is natural that $E_{\mathrm{VMC}}$ obtained by our HF$+$VMC calculations with $S_{\mathrm{ee}}$ and $S_{\mathrm{een}}$ lie between VMC total energies using very few and many variational parameters in Ref.~\cite{unreweighted}, as shown in Table~\ref{tab:tote_bene}.

Table~\ref{tab:oneshot_bene} presents the comparison between one-shot and self-consistent calculation, in terms of the calculated total energy, $E_{\mathrm{VMC}}$, and IP.
For calculation using $S_{\mathrm{ee}}$, self-consistent TC$+$VMC calculations are always superior to HF$+$VMC and one-shot TC$+$VMC.
On the other hand, calculation using $S_{\mathrm{een}}$, self-consistent TC$+$VMC can inferior to them. 
Possible causes of this behavior might be insufficiency of the degrees of freedom in our $S_{\mathrm{een}}$ Jastrow parameters, or the convergence problem as mentioned in the previous paragraph. In any case, our result show that one-shot TC$+$VMC can be a good alternative way for these cases, in terms of computational cost and the systematic accuracy. This problem might be alleviated when one uses the Jastrow function with a larger degree of freedom.

\begin{table*}
\caption{Total energy (Ht.) and the ratio of the correlation energy retrieved (\%) for beryllium and neon atoms. For evaluating the latter quantity, the HF energy calculated by us, $E_{\mathrm{HF}}=-14.5731(1)$ Ht. for Be and $-128.548(7)$ Ht. for Ne, and the exact total energy, $E_{\mathrm{exact}}=-14.66736$ Ht. for Be and $-128.939$ Ht. for Ne, estimated in Ref.~\onlinecite{He_totEex}, were used. $a=1.5$ Bohr was used in the Jastrow factor. The statistical error for our calculation results is not shown in this table because it is much less than the expected basis-set error $\sim 0.1$ mHt. We note that the variational principle breaks for $E_{\mathrm{TC}}$ and $E_{\mathrm{BITC}}$, which can result in the ratio of the correlation energy retrieved less than 0 \% or larger than 100 \%. For VMC calculations in Ref.~\cite{unreweighted}, a detail of variational parameters is shown in that reference.}
\vspace*{2mm}
\label{tab:tote_bene}
\begin{tabular}{c c c c c c c c c}
\hline
Atom & Method & Jastrow& $E_{\mathrm{VMC}}$ & (\%) & $E_{\mathrm{TC}}$ & (\%) & $E_{\mathrm{BITC}}$ & (\%)\\
 \hline \hline
Be & HF$+$VMC & $S_{\mathrm{ee}}$ & $-14.6072$ & $36.2$ & - & - & - & - \\
  & & $S_{\mathrm{een}}$ & $-14.6403$ & $71.3$ & - & - & - & - \\
  & TC$+$VMC & $S_{\mathrm{ee}}$ & $-14.6261$ & $56.3$ & $-14.6469$ & $78.3$ & $-14.6202$ & $50.0$\\
  & & $S_{\mathrm{een}}$ & $-14.6431$ & $74.3$ & $-14.6497$ & $81.3$ & $-14.6058$ & $34.7$\\
 \hline
 Ne & HF$+$VMC & $S_{\mathrm{ee}}$ & $-128.624$ & $19.3$ & - & - & - & - \\
  & & $S_{\mathrm{een}}$ & $-128.832$ & $72.6$ & - & - & - & - \\
  & TC$+$VMC & $S_{\mathrm{ee}}$ & $-128.745$ & $50.3$ & $-128.999$ & $115.4$ & $-128.651$ & $26.4$\\
  & & $S_{\mathrm{een}}$ & $-128.825$ & $70.8$ & $-129.077$ & $135.2$ & $-127.149$ & $-357.8$\\
  & \multicolumn{2}{c}{VMC with 1 linear optimizable parameter~\cite{unreweighted}} & $-128.6201(3)$ & $18.4$ & - & - & - & -\\
    & \multicolumn{2}{c}{VMC with 72 linear optimizable parameters~\cite{unreweighted}} & $-128.89752(7)$ & $89.4$ & - & - & - & -\\
 \hline
 \end{tabular}
 \end{table*}
 
\begin{table*}
\caption{Total energy (Ht.), the ratio of the correlation energy retrieved (\%), and the ionization potential (IP) (Ht.) estimated from the highest occupied orbital energy for each method, for beryllium and neon atoms. IP was evaluated by HF, TC, or BITC calculations (i.e., not by VMC). Calculated results for HF$+$VMC and self-consistent TC$+$VMC are taken from Table~\ref{tab:tote_bene}. The statistical error for our calculation results is not shown in this table because it is much less than the expected basis-set error $\sim 0.1$ mHt. As noted in Table~\ref{tab:notation}, BITC calculation in self-consistent TC$+$VMC was performed only once after TC$+$VMC calculation, i.e., using the Jastrow parameters determined by TC$+$VMC calculation.}
\vspace*{2mm}
\label{tab:oneshot_bene}
\begin{tabular}{c c c c c c c}
\hline
Atom & Jastrow & Method  &  \ \  $E_{\mathrm{VMC}}$  \ \  & \ \  (\%) \ \ & \ \ \ \ IP\ \ \ \ & \ \ \ \ IP (BITC)\ \ \ \ \\
 \hline \hline
Be & $S_{\mathrm{ee}}$ & HF$+$VMC & $-14.6072$ & $36.2$ & 0.3093 & -\\
 & & TC$+$VMC (one-shot)  & $-14.6189$ & $48.6$ & 0.3085 & 0.3104\\
 & & TC$+$VMC (self-consistent)  & $-14.6261$ & $56.3$ & 0.3064 & 0.3120\\
  \cline{2-7}
 & $S_{\mathrm{een}}$ & HF$+$VMC &  $-14.6403$ & $71.3$ & 0.3093 & -\\
 & & TC$+$VMC (one-shot) &  $-14.6449$ & $76.2$ & 0.2960 & 0.3017\\
 & & TC$+$VMC (self-consistent) &  $-14.6431$ & $74.3$  & 0.2769 & 0.2897\\
   \cline{5-7}
   & & & & Expt.\cite{Expt_IP} & \multicolumn{2}{c}{0.343} \\
    \hline
 Ne & $S_{\mathrm{ee}}$ & HF$+$VMC & $-128.624$ & $19.3$ & 0.851 & -\\
 & & TC$+$VMC (one-shot)  & $-128.679$ & $33.4$ & 0.913 & 0.856\\
   & & TC$+$VMC (self-consistent)  & $-128.745$ & $50.3$ & 0.797 & 0.858\\
  \cline{2-7}
 & $S_{\mathrm{een}}$ & HF$+$VMC &  $-128.832$ & $72.6$ & 0.851 & -\\
   & & TC$+$VMC (one-shot) &  $-128.854$ & $78.3$ & 0.765 & 0.833\\
      & & TC$+$VMC (self-consistent) &  $-128.825$ & $70.8$  & 0.546 & 0.816\\
  \cline{5-7}
   & & & & Expt.\cite{Expt_IP} & \multicolumn{2}{c}{0.792} \\
 \hline
 \end{tabular}
 \end{table*}
 
\subsection{TC$+$VMC (energy minimization) with one-body Jastrow terms\label{sec:one_body_emin}}
 
 As mentioned in Sec.~\ref{sec:TCVMC}, we found that an alternate repetition of TC and VMC calculations does not reach convergence when the energy is minimized in VMC calculations.
Nevertheless, TC$+$VMC (energy minimization) can reach convergence in some cases by including one-body terms in the Jastrow function as we shall see in this section.

For VMC calculations in this section, we minimized the total energy with including the cuspless one-body Jastrow terms, $(i,j,k)=(0,j,0)$ and $(0,0,k)$ with $2\leq j,k \leq 4$ in Eq.~(\ref{eq:Jastrow_pade}), in addition to $S_{\mathrm{ee}}$ or $S_{\mathrm{een}}$, as described in Sec.~\ref{sec:jastrow}.
On the other hand, for TC calculations, we did not include these additional one-body terms because these degrees of freedom can be considered through the orbital optimization. Since we considered the convergence of TC$+$VMC was achieved when the VMC energy is not further lowered in VMC optimization starting from $c_{0j0}=c_{00k}=0$ (i.e., no one-body terms), calculation results after convergence shown in this section do not depend on coefficients of the one-body Jastrow terms.

Because we did not succeed in reaching convergence for Ne, we present calculation results for He and Be atoms in Table~\ref{tab:emin}.
Overall, these results exhibit a consistent behavior with those obtained in TC$+$VMC calculations based on variance minimization.
For He, the ratios of the correlation energy retrieved evaluated with $E_{\mathrm{VMC}}$ are 91.9\% and 99.0\% for $S_{\mathrm{ee}}$ and $S_{\mathrm{een}}$, respectively, as is consistent with those obtained by TC$+$VMC calculations based on variance minimization: 90.9\% and 98.4\% for $S_{\mathrm{ee}}$ and $S_{\mathrm{een}}$ (see Table~\ref{tab:tote}), respectively.
We also see the same tendency for Tables~\ref{tab:tote} (variance minimization) and \ref{tab:emin} (energy minimization with one-body Jastrow terms) that $E_{\mathrm{TC}}$ tends to become lower than $E_{\mathrm{VMC}}$ and vice versa for $E_{\mathrm{BITC}}$.
Energy minimization tends to give a lower energy than variance minimization, which is also known in QMC calculations without TC orbital optimization.
For Be, the ratios of the correlation energy retrieved evaluated with $E_{\mathrm{VMC}}$ are 68.2\% and 79.7\% for $S_{\mathrm{ee}}$ and $S_{\mathrm{een}}$, respectively, as is consistent with those obtained in TC$+$VMC calculations based on variance minimization: 56.3\% and 74.3\% for $S_{\mathrm{ee}}$ and $S_{\mathrm{een}}$ (see Table~\ref{tab:tote_bene}), respectively.
It is problematic that $E_{\mathrm{TC}}$ and $E_{\mathrm{BITC}}$ are much different from $E_{\mathrm{VMC}}$. Increasing the number of the Jastrow parameters might alleviate it because these three estimates of the total energy coincide for the exact eigenstate.

\begin{table*}
\caption{Total energy (Ht.) and the ratio of the correlation energy retrieved (\%) for helium and beryllium atoms, calculated using VMC (energy minimization) including the one-body Jastrow functions. For evaluating the ratio of the correlation energy retrieved, the HF energy calculated by us, $E_{\mathrm{HF}}=-2.8617(3)$ Ht. for He and $E_{\mathrm{HF}}=-14.5731(1)$ Ht. for Be, and the exact total energy, $E_{\mathrm{exact}}=-2.90372$ Ht. for He and $E_{\mathrm{exact}}=-14.66736$ Ht. for Be, estimated in Ref.~\onlinecite{He_totEex}, were used. $a=1.5$ Bohr was used in the Jastrow factor. The statistical error for our calculation results is not shown in this table because it is much less than the expected basis-set error $\sim 0.1$ mHt. We note that the variational principle breaks for $E_{\mathrm{TC}}$ and $E_{\mathrm{BITC}}$, which can result in the ratio of the correlation energy retrieved less than 0 \% or larger than 100 \%.}
\vspace*{2mm}
\label{tab:emin}
\begin{tabular}{c c c c c c c c c}
\hline
Atom & Method & Jastrow& $E_{\mathrm{VMC}}$ & (\%) & $E_{\mathrm{TC}}$ & (\%) & $E_{\mathrm{BITC}}$ & (\%)\\
 \hline \hline
  He & TC$+$VMC & $S_{\mathrm{ee}}$ & $-2.9003$ & $91.9$ & $-2.9175$ & $132.9$ & $-2.9065$ & $106.7$\\
  & & $S_{\mathrm{een}}$ & $-2.9033$ & $99.0$ & $-2.9037$ & $99.9$ & $-2.9024$ & $96.7$\\
    \hline
Be  & TC$+$VMC & $S_{\mathrm{ee}}$ & $-14.6374$ & $68.2$ & $-14.7209$ & $156.7$ & $-14.5598$ & $-14.1$\\
  & & $S_{\mathrm{een}}$ & $-14.6482$ & $79.7$ & $-14.6678$ & $100.4$ & $-14.5778$ & $5.0$\\
 \hline
 \end{tabular}
 \end{table*}
 
\section{Conclusion\label{sec:Con}}

In this study, we have investigated how TC$+$VMC calculation works for small atoms.
Important findings are summarized as follows.
\begin{enumerate}
\item TC$+$VMC calculation can successfully reach a self-consistent solution by adopting variance minimization in VMC, while energy minimization in some cases works well by including one-body Jastrow terms.
\item The total energy evaluated by VMC ($E_{\mathrm{VMC}}$) is in many cases systematically improved by using better Jastrow functions, and the expectation value of the TC Hamiltonian, $E_{\mathrm{TC}}$, gets closer to $E_{\mathrm{VMC}}$, accordingly.
However, it is suggested that $E_{\mathrm{VMC}}$ is a better estimate of the total energy than $E_{\mathrm{TC}}$ and $E_{\mathrm{BITC}}$, when the number of Jastrow parameters is not sufficient.
\item One can partially receive the benefit of the orbital optimization even by one-shot TC$+$VMC, where the Jastrow parameters are optimized at the HF$+$VMC level.
One-shot TC$+$VMC calculation can be a good compromise in complex systems.
\end{enumerate}

Our study provides important knowledge for optimizing many-body wave function including the Jastrow correlation factor, which would be of great help for development of highly accurate electronic structure calculation.

\section*{ACKNOWLEDGMENTS} 

This study was supported by Grant-in-Aid for young scientists, Grant Number JP18K13470, from the Japan Society for the Promotion of Science, Japan, and JST FOREST Program, Grant Number JPMJFR212P.
We thank Prof. Shinji Tsuneyuki for fruitful discussion.

\section*{Appendix A: Orthonormality of the basis set $f_n^{l_i}$}
Using a formula
\begin{equation}
\int_0^{\infty} x^{a} e^{-x} L_m^{(a)}(x) L_n^{(a)}(x) \mathrm{d}x = \frac{(n+a)!}{n!}\delta_{m,n},\label{eq:orthonorm_aL}
\end{equation}
one can verify that $f_n^{l_i}$ is orthonormalized as follows:
\begin{widetext}
\begin{align}
\int_0^{\infty} f_n^{l_i}(r) f_m^{l_i}(r) \mathrm{d}r 
&=\sqrt{\frac{m!n!}{(m+2l_i+2)!(n+2l_i+2)!}} \int_0^{\infty} (2 \alpha r_i)^{2l_i+2}  e^{-2\alpha r} L_m^{2l_i+2}(2\alpha r)  L_n^{2l_i+2}(2\alpha r)   \mathrm{d}(2\alpha r) \\
&=\sqrt{\frac{m!n!}{(m+2l_i+2)!(n+2l_i+2)!}} \frac{(n+2l_i+2)!}{n!}\delta_{m,n} \\
&=\delta_{m,n}.
\end{align}
\end{widetext}

\section*{Appendix B: Calculation of Eq.~(\ref{eq:1body})}

For simplicity, we omit the normalization constant in $f_n^{l_i}$ here.
To say, we evaluate the integral for $\tilde{f}_n^{l_i} = r^{l_i+1} L_n^{(2l_i+2)}(2\alpha r) e^{-\alpha r}$ instead of that for $f_n^{l_i}$.
We only consider the $m\geq n$ case, the other case of which ($m<n$) is readily derived by Hermiticity.
The first term in Eq.~(\ref{eq:1body}) is written as follows:
 \begin{widetext}
\begin{align}
-\frac{1}{2} \int_0^{\infty}\mathrm{d}r\  \tilde{f}_m^{l_i}(r) \frac{\mathrm{d}^2}{\mathrm{d}r^2}&\tilde{f}_n^{l_i}(r)\\
= -\frac{1}{2}\int_0^{\infty}\mathrm{d}r\  \tilde{f}_m^{l_i}(r) &\bigg[ \left( \frac{\mathrm{d}^2}{\mathrm{d}r^2}r^{l_i+1} \right) L_n^{(2l_i+2)}(2\alpha r) e^{-\alpha r}
+ r^{l_i+1} \left(  \frac{\mathrm{d}^2}{\mathrm{d}r^2}L_n^{(2l_i+2)}(2\alpha r) \right)  e^{-\alpha r} \label{eq:app2}\\
&+ r^{l_i+1} L_n^{(2l_i+2)}(2\alpha r) \left(  \frac{\mathrm{d}^2}{\mathrm{d}r^2}e^{-\alpha r} \right)
+2 \left( \frac{\mathrm{d}}{\mathrm{d}r}r^{l_i+1} \right) \left( \frac{\mathrm{d}}{\mathrm{d}r}L_n^{(2l_i+2)}(2\alpha r) \right) e^{-\alpha r} \notag \\
&+2 \left( \frac{\mathrm{d}}{\mathrm{d}r}r^{l_i+1} \right) L_n^{(2l_i+2)}(2\alpha r)  \left( \frac{\mathrm{d}}{\mathrm{d}r} e^{-\alpha r} \right) 
+2 r^{l_i+1} \left( \frac{\mathrm{d}}{\mathrm{d}r}L_n^{(2l_i+2)}(2\alpha r) \right)  \left( \frac{\mathrm{d}}{\mathrm{d}r} e^{-\alpha r} \right) \bigg] .\notag
\end{align}
\end{widetext}
The first term in Eq.~(\ref{eq:app2}) is rewritten as $\int_0^{\infty}\mathrm{d}r\  \tilde{f}_m^{l_i}(r) \frac{-l_i(l_i+1)}{2r^2} \tilde{f}_n^{l_i}(r)$, which cancels out the second term in Eq.~(\ref{eq:1body}).
The second and sixth terms in Eq.~(\ref{eq:app2}) are zero because of the orthonormalization condition (Eq.~(\ref{eq:orthonorm_aL})) and the fact that any $l$-th polynomial can be represented as linear combination of $L_i^{(a)}$ with $i\leq l$. The third term in Eq.~(\ref{eq:app2}) gives the first term in Eq.~(\ref{eq:1body_2}).
For calculating the fourth term, we note
\begin{widetext}
\begin{align}
\left( \frac{\mathrm{d}}{\mathrm{d}r}r^{l_i+1} \right) \left( \frac{\mathrm{d}}{\mathrm{d}r}L_n^{(2l_i+2)}(2\alpha r) \right)
&= r^{l_i+1} \times ((n-2)\mathrm{-th \ order\ polynomial}) - 2\alpha(l_i+1) r^{l_i} \left(
	\begin{array}{c}
	n+2l_i+2 \\
	n-1 \\
	\end{array}
\right) \\
&= r^{l_i+1} \times ((n-1)\mathrm{-th \ order\ polynomial}) - 2\alpha(l_i+1) r^{l_i} \frac{n}{2l_i+3} L_n^{(2l_i+2)}(2\alpha r),\label{eq:fourth} 
\end{align}
\end{widetext}
which is derived using the coefficients of the zeroth- and first-order terms in the associated Laguerre polynomial.
The first term in Eq.~(\ref{eq:fourth}) makes no contribution to Eq.~(\ref{eq:app2}) because of the orthogonality of the associated Laguerre polynomials and $m>n-1$.
The second term in Eq.~(\ref{eq:fourth}) yields
\begin{equation}
2\alpha \frac{(l_i+1)n}{2l_i+3} \int_0^{\infty}\mathrm{d}r\  \tilde{f}_m^{l_i}(r) \frac{1}{r} \tilde{f}_n^{l_i}(r).
\end{equation}
Note that $n$ in the coefficient of this integral should be replaced with $\mathrm{min}(m,n)$ when one considers not only the $m\geq n$ case but also $m<n$.
The fifth term in Eq.~(\ref{eq:app2}) gives $\int_0^{\infty}\mathrm{d}r\  \tilde{f}_m^{l_i}(r) \frac{\alpha(l_i+1)}{r} \tilde{f}_n^{l_i}(r)$.
By summing up all of them, we can obtain Eq.~(\ref{eq:1body_2}) (see also Appendix in \cite{Laguerre1}).

\section*{Appendix C: Derivation of Eq.~(\ref{eq:Ylmderiv})}
Spherical harmonics is defined as follows:
\begin{equation}
Y_{l,m}(\Omega) = (-1)^{\frac{m+|m|}{2}}\sqrt{\frac{2l+1}{4\pi}\frac{(l-|m|)!}{(l+|m|)!}}P_l^{|m|}(\cos \theta) e^{im\varphi},
\end{equation}
where $P_{\nu}^{\mu}(t)$ are associated Legendre polynomials satisfying the following formula,
\begin{equation}
\sin^2 \theta \frac{\mathrm{d}}{\mathrm{d}t} P_{\nu}^{\mu}(t) = \sin \theta P_{\nu}^{\mu+1}(t) - \mu \cos \theta P_{\nu}^{\mu}(t).
\end{equation}
By using these equations, we can immediately derive Eq.~(\ref{eq:Ylmderiv}).

\section*{Appendix D: Derivatives of the Jastrow function used in this study}
While the analytical derivative of the Jastrow function, Eq.~(\ref{eq:Jastrow_pade}), is straight-forward, we present mathematical expressions for reference, which can be helpful in implementing the TC method. A simple calculation yields
\begin{gather}
\mathbf{\nabla}_1 u(x_1,x_2) = \sum_{(i,j,k)\in S} c^{\sigma_1\sigma_2}_{ijk} \bar{r}^{i-1}_{12} \bar{r}_1^{j-1} \bar{r}_2^k \notag \\
\times \left( \frac{\mathbf{r}_{12}}{r_{12}}\frac{ia_{12}}{(r_{12}-a_{12})^2} \bar{r}_1 +  \frac{\mathbf{r}_{1}}{r_{1}}\frac{ja}{(r_{1}-a)^2} \bar{r}_{12}\right),
\end{gather}
and
\begin{gather}
\nabla_1^2 u(x_1,x_2) = \sum_{(i,j,k)\in S} c^{\sigma_1\sigma_2}_{ijk} \bar{r}^{i-2}_{12} \bar{r}_1^{j-2} \bar{r}_2^k \notag \\
\times \bigg( \frac{(i+1)ia_{12}^2}{(r_{12}+a_{12})^4} \bar{r}_1^2 + \frac{(j+1)ja^2}{(r_{1}+a)^4} \bar{r}_{12}^2 \\
+ \frac{\mathbf{r}_{12}\cdot\mathbf{r}_1}{r_{12}r_1}\frac{2ija_{12}a}{(r_{12}+a_{12})^2(r_1+a)^2} \bigg),
\end{gather}
by which we can readily write down the TC effective potentials represented with $\mathbf{\nabla}u$ and $\nabla^2 u$.
We note that a sign of the coefficients $c^{\sigma_1\sigma_2}_{ijk}$ should be reversed when one defines the Jastrow factor as
$F=\mathrm{exp}(\sum_{i,j(\neq i)}^N u(x_i,x_j))$, unlike our notation shown in Eq.~(\ref{eq:Jastrowfactor}).


\begin{thebibliography}{999}

\bibitem{cusp} T. Kato, Commun. Pure Appl. Math. \textbf{10}, 151 (1957).
\bibitem{cusp2} R. T. Pack and W. B. Brown, J. Chem. Phys. \textbf{45}, 556 (1966).

\bibitem{Hyl1} E. A. Hylleraas, Z. Phys. \textbf{48}, 469 (1928).
\bibitem{Hyl2} E. A. Hylleraas, Z. Phys. \textbf{54}, 347 (1929).
\bibitem{Hyl3} E. A. Hylleraas, Z. Phys. \textbf{65}, 209 (1930).

\bibitem{R12} W. Kutzelnigg, Theor. Chim. Acta \textbf{68}, 445 (1985).
\bibitem{F12} S. Ten-no, Chem. Phys. Lett. \textbf{398}, 56 (2004).

\bibitem{QMC} R. J. Needs, M. D. Towler, N. D. Drummond, and P. L{\'o}pez R{\'ios}, J. Phys.: Condens. Matter {\bf 22}, 023201 (2010).
\bibitem{QMCsolids} W. M. C. Foulkes, L. Mitas, R. J. Needs, and G. Rajagopal, Rev. Mod. Phys. {\bf 73}, 33 (2001).

\bibitem{BoysHandy} S. F. Boys and N. C. Handy, Proc. R. Soc. London Ser. A \textbf{309}, 209 (1969); {\it ibid.} \textbf{310}, 43 (1969); {\it ibid.} \textbf{310}, 63 (1969); {\it ibid.} \textbf{311}, 309 (1969).
\bibitem{Handy} N. C. Handy, Mol. Phys. \textbf{21}, 817 (1971).

\bibitem{Ten-no1} S. Ten-no, Chem. Phys. Lett. \textbf{330}, 169 (2000); {\it ibid.} 175 (2000). 
\bibitem{Ten-no2} O. Hino, Y. Tanimura, S. Ten-no, J. Chem. Phys. \textbf{115}, 7865 (2001). 
\bibitem{Umezawa} N. Umezawa and S. Tsuneyuki, J. Chem. Phys. \textbf{119}, 10015 (2003). 

\bibitem{Ten-no3} O. Hino, Y. Tanimura, S. Ten-no, Chem. Phys. Lett. \textbf{353}, 317 (2002). 
\bibitem{TCCC2021} T. Schraivogel, A. J. Cohen, A. Alavi, and D. Kats, J. Chem. Phys. {\bf 155}, 191101 (2021). 

\bibitem{Umezawa_CIS} N. Umezawa and S. Tsuneyuki, J. Chem. Phys. {\bf 121}, 7070 (2004). 
\bibitem{LuoVTC} H. Luo, J. Chem. Phys. \textbf{133}, 154109 (2010). 
\bibitem{Luo_multiconf} H. Luo, J. Chem. Phys. {\bf 135}, 024109 (2011). 
\bibitem{Giner_He} E. Giner, J. Chem Phys. {\bf 154}, 084119 (2021). 
\bibitem{SCI} A. Ammar, A. Scemama, and E. Giner, J. Chem. Phys. {\bf 157}, 134107 (2022).

\bibitem{CanonicalTC} T. Yanai and T. Shiozaki, J. Chem. Phys. \textbf{136}, 084107 (2012).
\bibitem{CanonicalTC_qCCSD} M. Motta, T. P. Gujarati, J. E. Rice, A. Kumar, C. Masteran, J. A. Latone, E. Lee, E. F. Valeev, and T. Y. Takeshita, Phys. Chem. Chem. Phys. {\bf 22}, 24270 (2020). 

\bibitem{FCI_canoTC_DMRG} S. Sharma, T. Yanai, G. H. Booth, C. J. Umrigar, and G. K.-L. Chan, J. Chem. Phys. {\bf 140}, 104112 (2014). 
\bibitem{FCI_canoTC} J. A. F. Kersten, G. H. Booth, and A. Alavi, J. Chem. Phys. {\bf 145}, 054117 (2016). 

\bibitem{FCI_Bedimer} K. Guther, A. J. Cohen, H. Luo, and A. Alavi, J. Chem. Phys. {\bf 155}, 011102 (2021). 
\bibitem{FCI_largeCI} A. Ammar, E. Giner, and A. Scemama, J. Chem. Theory Comput. {\bf 18}, 5325 (2022).
\bibitem{FCI_Hubbard} W. Dobrautz, H. Luo, and A. Alavi, Phys. Rev. B {\bf 99}, 075119 (2019).
\bibitem{FCI_TC_elgas} H. Luo and A. Alavi, J. Chem. Theory Comput. {\bf 14}, 1403 (2018). 

\bibitem{FCI_1dgas} P. Jeszenszki, H. Luo, A. Alavi, and J. Brand, Phys. Rev. A {\bf 98}, 053627 (2018). 

\bibitem{FCI_cold} P. Jeszenszki, U. Ebling, H. Luo, A. Alavi, and J. Brand, Phys. Rev. Res. {\bf 2}, 043270 (2020). 

\bibitem{McArdle} S. McArdle and D. P. Tew, arXiv:2006.11181.
\bibitem{quantum_simulation} A. Kumar, A. Asthana, C. Masteran, E. F. Valeev, Y. Zhang, L. Cincio, S. Tretiak, and P. A. Dub, J. Chem. Theory Comput. {\bf 18}, 5312 (2022).

\bibitem{Sakuma} R. Sakuma and S. Tsuneyuki, J. Phys. Soc. Jpn. \textbf{75}, 103705 (2006).
\bibitem{TCaccel} M. Ochi, K. Sodeyama, R. Sakuma, and S. Tsuneyuki, J. Chem. Phys. \textbf{136}, 094108 (2012).
\bibitem{TCjfo} M. Ochi, K. Sodeyama, and S. Tsuneyuki, J. Chem. Phys. \textbf{140}, 074112 (2014).
\bibitem{TCPW} M. Ochi, Y. Yamamoto, R. Arita, and S. Tsuneyuki, J. Chem. Phys. \textbf{144}, 104109 (2016).
\bibitem{TCZnO} M. Ochi, R. Arita, and S. Tsuneyuki, Phys. Rev. Lett. \textbf{118}, 026402 (2017).
\bibitem{TCPP} M. Ochi, Comput. Phys. Commun. \textbf{287}, 108687 (2023).
\bibitem{TCCIS} M. Ochi and S. Tsuneyuki, J. Chem. Theory Comput. \textbf{10}, 4098 (2014).
\bibitem{TCMP2} M. Ochi and S. Tsuneyuki, Chem. Phys. Lett. \textbf{621}, 177 (2015).

\bibitem{TCHubbard} S. Tsuneyuki, Prog. Theor. Phys. Suppl. \textbf{176}, 134 (2008).
\bibitem{TCDMRG} A. Baiardi and M. Reiher, J. Chem. Phys. {\bf 153}, 164115 (2020).
\bibitem{LieAlgebra} J. M. Wahlen-Strothman, C. A. Jim{\' e}nez-Hoyos, T. M. Henderson, and G. E. Scuseria, Phys. Rev. B \textbf{91}, 041114(R) (2015).

\bibitem{elgas_Armour} E. A. G. Armour, J. Phys. C: Solid State Phys. {\bf 13}, 343 (1980).
\bibitem{Umezawa_elgas} N. Umezawa and S. Tsuneyuki, Phys. Rev. B {\bf 69}, 165102 (2004).
\bibitem{Luo_elgas} H. Luo, J. Chem. Phys. {\bf 136}, 224111 (2012).
\bibitem{perturbation_elgas} H. Luo and A. Alavi, J. Chem. Phys. {\bf 157}, 074105 (2022).

\bibitem{Umezawa_beta} R. Prasad, N. Umezawa, D. Domin, R. Salomon-Ferrer, and W. A. Lester, Jr., J. Chem. Phys. \textbf{126}, 164109 (2007). 
\bibitem{LuoTC} H. Luo, W. Hackbusch, and H.-J. Flad, Mol. Phys. \textbf{108}, 425 (2010). 
\bibitem{TCatoms_HFJastrow} A. J. Cohen, H. Luo, K. Guther, W. Dobrautz, D. P. Tew, and A. Alavi, J. Chem. Phys. {\bf 151}, 061101 (2019). 
\bibitem{TCatoms_oneparam} W. Dobrautz, A. J. Cohen, A. Alavi, and E. Giner, J. Chem. Phys. {\bf 156}, 234108 (2022). 
\bibitem{TCatoms_2023} J. P. Haupt, S. M. Hosseini, P. L. R{\'i}os, W. Dobrautz, A. Cohen, and A. Alavi, J. Chem. Phys. {\bf 158}, 224105 (2023).

\bibitem{Umrigar_fullopt} J. Toulouse and C. J. Umrigar, J. Chem. Phys. {\bf 126}, 084102 (2007). 
\bibitem{Umrigar_fullopt2} J. Toulouse and C. J. Umrigar, J. Chem. Phys. {\bf 128}, 174101 (2008).


\bibitem{BohmPines} D. Bohm and D. Pines, Phys. Rev. \textbf{92}, 609 (1953).

\bibitem{Laguerre1} S. Hagstrom and H. Shull, J. Chem. Phys. \textbf{30}, 1314 (1959).
\bibitem{Laguerre2} Y. Hatano and S. Yamamoto, J. Comput. Chem. Jpn. Int. Ed. \textbf{2}, 2016-0003 (2016).

\bibitem{asympt} The asymptotic behavior of the Hartree-Fock orbitals is investigated in old literature~\cite{asympt_HF}, and the same discussion holds for the TC and BITC methods. This fact originates from the exchange terms in the SCF equation, which brings $e^{- \sqrt{-2\epsilon_{\mathrm{HO}}} r}$ asymptotic behavior for $r\to \infty$ (to say, the slowest decay among the occupied orbitals) to all the orbitals.

\bibitem{asympt_HF} N. C. Handy, M. T. Marron, and H. J. Silverstone, Phys. Rev. {\bf 180}, 45 (1969).

\bibitem{note_homo} When $\epsilon_{\mathrm{HO}}$ happens to be positive in some SCF loop, we instead substituted some negative value such as $\epsilon_{\mathrm{max}} = -0.05$ Ry for $\alpha=\sqrt{-2\epsilon_{\mathrm{max}}}$ to keep the decaying behavior of the basis functions in the $r\to \infty$ limit. Of course, this should not happen in the final SCF loop.

\bibitem{note_matrix} For the BITC method, the operator acting on the right orbital $\phi$ in the left-hand side of the SCF equation is Hermitian conjugate of that for the left orbital $\chi$. Thus, right and left orbitals are simultaneously obtained as the right and left eigentstates by diagonalizing the common matrix as discussed in Sec.~\ref{sec:TC}.

\bibitem{poly1} K. E. Schmidt and J. W. Moskowitz, J. Chem. Phys. {\bf 93}, 4172 (1990).
\bibitem{poly2} C. Filippi and C. J. Umrigar, J. Chem. Phys. {\bf 105}, 213 (1996).

\bibitem{Ten-nocusp} S. Ten-no, J. Chem. Phys. \textbf{121}, 117 (2004).
\bibitem{cuspUmrigar} C.-J. Huang, C. Filippi, C. J. Umrigar, J. Chem. Phys. {\bf 108}, 8838 (1998).

\bibitem{CASINO} R. J. Needs, M. D. Towler, N. D. Drummond, P. L{\'o}pez R{\'i}os, and J. R. Trail, J. Chem. Phys. {\bf 152}, 154106 (2020).

\bibitem{grid1} A. D. Becke, J. Chem. Phys. {\bf 88}, 2547 (1988).
\bibitem{grid2} O. Treutler and R. Ahlrichs, J. Chem. Phys. {\bf 102}, 346 (1995).

\bibitem{unreweighted} N. D. Drummond and R. J. Needs, Phys. Rev. B {\bf 72}, 085124 (2005).

\bibitem{HFCBS} K. Szalewicz and H. J. Monkhorst, J. Chem. Phys. {\bf 75}, 5785 (1981).
\bibitem{He_totEex} E. R. Davidson, S. A. Hagstrom, S. J. Chakravorty, V. M. Umar, and C. F. Fischer, Phys. Rev. A {\bf 44}, 7071 (1991).

\bibitem{He_VMC} N. D. Drummond, M. D. Towler, and R. J. Needs, Phys. Rev. B {\bf 70}, 235119 (2004).

\bibitem{Expt_IP} A. A. Radzig and B. M. Smirnov, ``Reference Data on Atoms, Molecules, and Ions'', Springer, Berlin (1985).

\end{thebibliography}
\end{document}